\documentclass[journal=jctc,manuscript=article]{achemso}

\usepackage[version=3]{mhchem} 
\usepackage{graphicx}
\usepackage{amsmath}
\usepackage{amssymb}
\usepackage{dcolumn}
\usepackage{color}
\usepackage[normalem]{ulem}
\usepackage{chemformula}
\usepackage{float}

\renewcommand{\Ref}[1]{Ref.~\citenum{#1}}
\DeclareMathOperator{\cosech}{cosech}

\SectionNumbersOn

\author{Wei Fang}
  
\author{Pierre Winter}

\author{Jeremy O. Richardson}
\email{jeremy.richardson@phys.chem.ethz.ch}
\affiliation{Laboratory of Physical Chemistry, ETH Zürich, 8093 Zürich, Switzerland}

\title{Microcanonical Tunneling Rates from Density-of-States Instanton Theory}
\begin{document}

\makeatletter
\setlength\acs@tocentry@height{7.5cm}
\setlength\acs@tocentry@width{5.5cm}
\makeatother

\begin{tocentry}

\includegraphics[width=7cm]{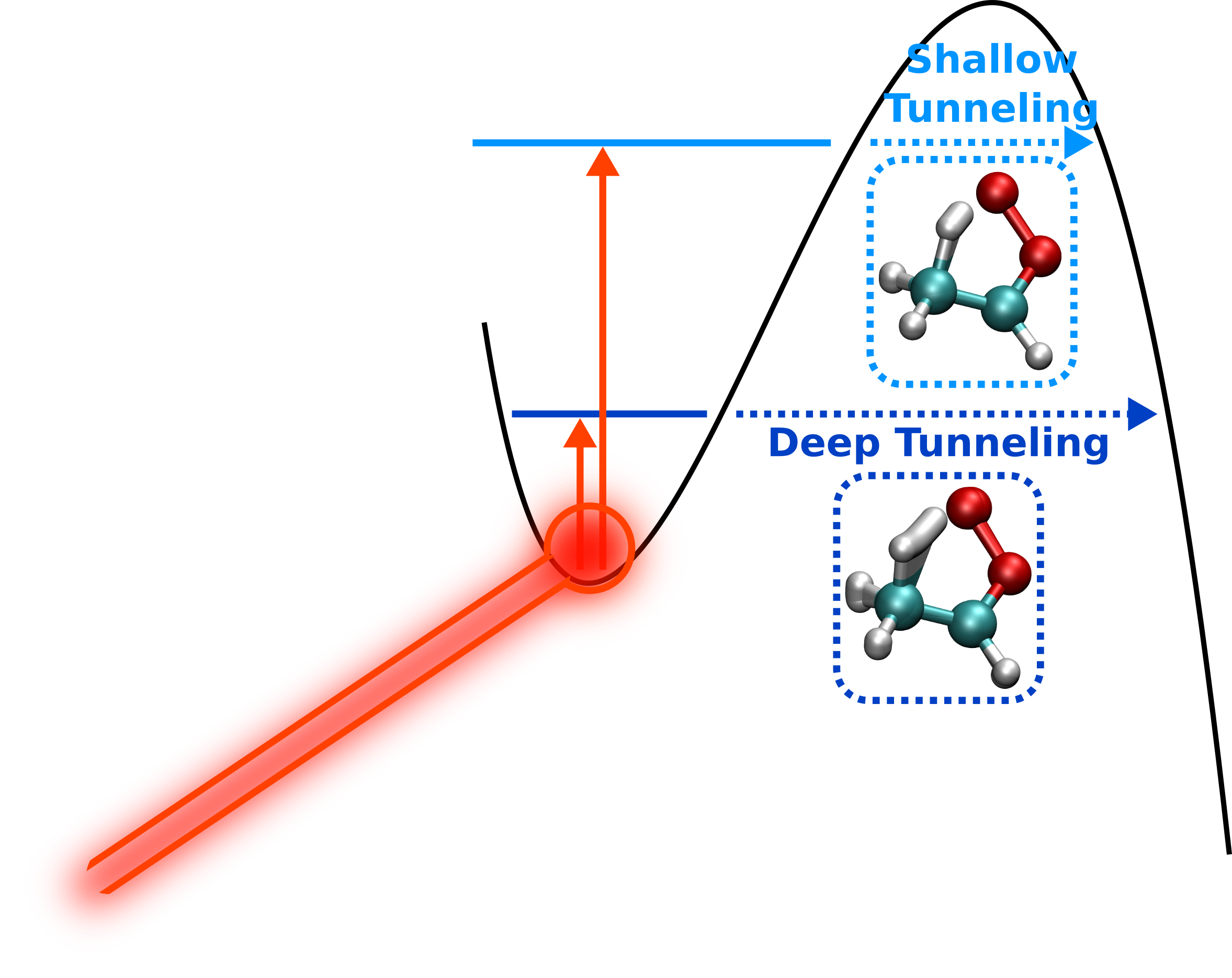}

\end{tocentry}

\begin{abstract}
Semiclassical instanton theory is a form of quantum transition-state theory which can be applied to computing thermal reaction rates for complex molecular systems including quantum tunneling effects.
There have been a number of attempts to extend the theory to treat microcanonical rates.
However, the previous formulations are either computationally unfeasible for large systems due to an explicit sum over states
or they involve extra approximations which make them less reliable.
We propose a robust and practical microcanonical formulation called density-of-states instanton theory, which 
avoids the sum over states altogether.
In line with the semiclassical approximations inherent to the instanton approach, we employ the stationary-phase approximation to the inverse Laplace transform to obtain the densities of states.
This can be evaluated using only post-processing of the data available from a small set of instanton calculations, such that our approach remains computationally efficient.
We show that the new formulation predicts results that agree well with 
quantum scattering theory for an atom-diatom reaction and
with experiments for a photoexcited unimolecular hydrogen transfer in a Criegee intermediate.
When the thermal rate is evaluated from a Boltzmann average over our new microcanonical formalism,
it can overcome some problems of conventional instanton theory.  
In particular, it predicts a smooth transition at the crossover temperature and is able to describe bimolecular reactions with pre-reactive complexes such as \ch{CH3OH} + OH.
\end{abstract}

\section{Introduction}
Many reactions of interest cannot be accurately modeled with elementary thermal rate theories.
The simplest examples are unimolecular reactions initiated by energy-specific excitations. \cite{Quack_IR_rev_1987,Quack_rate_1974}
However, cases are also provided by reaction schemes which proceed via one or more intermediates or pre-reactive complexes. \cite{multiwell2000, MESMER, POLYRATE, ME_Klippenstein, microcanonical_jasper, Blitz2006}
In a low-pressure environment the reaction will not thermalize at each step
and thus thermal rate theories such as transition-state theory (TST) would not be valid. 
However, assuming that the reaction is statistically well-behaved, the total rate constant can be computed by averaging over an appropriate distribution of reaction energies using a microcanonical framework for each step.
The development of a robust method for computing microcanonical rates is thus important not only for predicting energy-dependent unimolecular rate constants, but is also the key to obtaining the total rate constant in complex reaction schemes.

One of the most widely used microcanonical rate theories is Rice–Ramsperger–Kassel–Marcus (RRKM) theory. \cite{RRKM1, RRKM2}
One of the main assumptions in RRKM theory is that the time scale of intramolecular energy redistribution is much faster than the time scale of the reaction.
This assumption allows one to employ a statistical approach to predict the reaction probability without requiring knowledge of the full dynamics of the system, which can be prohibitively expensive to compute.
Many studies have found that RRKM theory gives qualitative or even quantitative agreement with experiments and can be a robust model for predicting chemical reaction rates. \cite{Wang_Propyne_1999,Plane_Experimental_2019}
However, because it is a classical theory, it is not valid for energies below the reaction barrier and is thus limited to describing high-energy excitations.
In reality, however, the reaction may still proceed at low energies by penetrating the barrier via a quantum tunneling mechanism. 
Neglecting these nuclear quantum effects can lead to a drastic under-prediction of the rate constant and it is thus necessary to incorporate them in our kinetic models. \cite{BellBook, Benderskii, Ley2012tunnelling, Schreiner2011C2H4O, Carpenter2011tunnelling, Meisner2016review} 

Simple methods for adding one-dimensional tunneling corrections to RRKM theory have been previously developed. \cite{Miller1979unimolecular,VTSTtunnel}
These approaches rely on the assumption that the dynamics along a predetermined reaction coordinate are separable from the other degrees of freedom of the system.
This is clearly an approximation which can perform well in some cases, \cite{Watson2018Criegee} but will break down in others, especially when strong corner-cutting effects occur. \cite{Marcus1977cornercut,hexamerprism}

Instanton theory provides a rigorous approach to go beyond the separable approximation by defining the unique optimal tunneling pathway (known as the ``instanton'') through a multidimensional potential-energy barrier. \cite{Perspective,Miller1975semiclassical} 
This theory can be viewed as a semiclassical approximation to the exact quantum thermal rate. \cite{Miller1975semiclassical,AdiabaticGreens,InstReview}
Using the ring-polymer instanton formalism, \cite{Andersson2009Hmethane,RPInst,Rommel2011locating,Perspective,InstReview} the theory is computationally feasible for calculating thermal reaction rate constants for large systems in full dimensionality, even when using \textit{ab initio} electronic-structure. \cite{dimersurf,hexamerprism,i-wat2,porphycene,HCH4,GPR,*Muonium,DaC,Rommel2012enzyme,Meisner2017water,Asgeirsson2018instanton,Kryvohuz2012instanton}
Unfortunately, the approximations of conventional thermal instanton theory break down for reactions near the crossover temperature, \cite{Zhang2014interpolation,McConnell2017instanton} as well as for those which tunnel through broad-top barriers \cite{broadtop} or proceed via pre-reactive complexes. \cite{Meisner2016water,Kastner2016Faraday}
A practical instanton theory based on the microcanonical framework is thus highly desired as it could in principle fix all of these issues. \cite{Faraday}
While a number of instanton-based microcanonical rate theories have been proposed in the past, some 
may behave unphysically due to a failure to describe separable modes correctly. \cite{Miller1975semiclassical,Faraday}
Although an improved formula has been suggested, \cite{Chapman1975rates} this is not simple to apply in practice to large molecular systems due to the necessity of explicitly summing over all the vibrationally excited states.
An alternative version makes the formula more practical by solving it analytically using 3rd or 4th-order Taylor approximations of the PES around the barrier top, \cite{Miller1990SCTST,Greene20161D} which works well at high energies, but will not always be able to correctly describe the instanton pathways involved in deep tunneling. \cite{Goel2018SCTST}

In this work, we propose a new method for computing microcanonical rates that is capable of describing both quantum mechanical deep tunneling and classical over-the-barrier transmission.
Our method is based on instanton theory such that it goes beyond the separable approximation and describes reactions with the optimal tunneling pathway. 
Unlike previous microcanonical instanton theories, 
our formulation avoids direct summation over all the excited states,
but instead integrates over the density of states obtained from an inverse Laplace transform (ILT).
The ILT can be computed with the stationary phase approximation (SPA), which appears to be accurate and numerically stable.
We demonstrate that our new formulation is computationally inexpensive (unlike the direct-sum approach) as it only weakly depends on the system size in the post-processing of data obtained from a handful of separate instanton optimizations.
Like the direct-sum approach, it also recovers the correct classical limit for transmission above the barrier and treats separable modes correctly.
We show how a thermal rate can be obtained from our new microcanonical framework and prove that it reduces to conventional thermal instanton rate theory under appropriate conditions.
Finally, we demonstrate examples of real-world applications of our method, in which it performs well even when methods based on the separable approximation break down.

In this paper, we first briefly review classical RRKM theory and previous tunneling corrections based on the separable approximation. 
We then introduce thermal instanton theory and previous microcanonical rate theories.
Finally we describe our new density-of-states instanton approach in detail and apply it to molecular systems in full dimensionality, including a unimolecular hydrogen transfer in \textit{syn-}\ch{CH3CHOO} initiated by photo-excitation and the bimolecular \ch{CH3OH + OH} reaction, which proceeds via a pre-reactive complex.

\section{Theory}
The microcanonical rate constant for a unimolecular reaction is defined by \cite{RRKM1,KineticsBook}
\begin{equation}
k(E)=\frac{1}{2\pi\hbar}\frac{N(E)}{\rho_\text{r}(E)},
\label{kMI}
\end{equation}
where $N(E)$ is the cumulative reaction probability for a molecular system in the center-of-mass frame with total rovibrational energy $E$.
The term in the denominator,
$\rho_\text{r}(E)$,
is the reactant density of states, which can be defined as the number of vibrational and rotational quantum states per unit energy.
For a bimolecular reaction, the same expression for the microcanonical rate is used, although the energy $E$ now also includes the collision energy (from the relative translational motion).

The unimolecular thermal reaction rate constant at temperature $T$ is given by a Boltzmann average over the microcanonical rate, and can be written as \cite{Miller1974QTST}
\begin{equation}
k_\text{uni}(\beta) = \frac{1}{Z_{\text{r}}(\beta)}
\int_0^{\infty} \mathrm{d}E~k(E) \rho_\text{r}(E) \mathrm{e}^{-\beta E} =
\frac{1}{2\pi\hbar}\frac{1}{Z_{\text{r}}(\beta)}\int_{\mathcal{E}_{\text{r}}}^{\infty}\text{d}E~N(E)\mathrm{e}^{-\beta E},
\label{kint}
\end{equation}
where here the parameter $\beta$ is fixed by the temperature through the relation $\beta\equiv\beta(T)=1/k_\text{B}T$, and $k_\text{B}$ is the Boltzmann constant. 
The energy scale has been chosen such that the minimum potential energy of the reactant is 0,
but in the final expression, we have explicitly indicated the lower limit of the energy integral as $\mathcal{E}_{\text{r}}$,
the reactant zero-point energy (ZPE)\@.
It is clear that no reaction can take place with a total energy below this value.
$Z_{\text{r}}(\beta)$ is the rovibrational reactant partition function and
is related to the density of states via the Laplace transform, $Z_{\text{r}}(\beta) = \int_0^{\infty}\text{d}E~\rho_\text{r}(E)\mathrm{e}^{-\beta E}$.
We note therefore that the reactant density of states can alternatively be defined via an ILT of the reactant partition function as described in Appendix~\ref{app:ILT}.
Equation~\eqref{kint} is only appropriate for unimolecular reactions, in which the overall translation of the system does not play a role.
For bimolecular (scattering) reactions, the thermal rate constant must be written $k_\text{bi}(\beta) = \frac{Z_\text{TS}^{\text{trans}}(\beta)}{Z_{\text{r}}^{\text{trans}}(\beta)}k_\text{uni}(\beta)$, where $Z_{\text{r}}^{\text{trans}}(\beta)$ and $Z_\text{TS}^{\text{trans}}(\beta)$ are the reactant and transition state (TS) translational partition functions respectively and again $\beta\equiv\beta(T)$.

From these definitions, one can see that obtaining a reliable approximation to $N(E)$ is the key not only for predicting microcanonical rates, but also for improving thermal rate theories.
The remainder of this paper briefly discusses previous approaches and introduces our new density-of-states instanton method for calculating $N(E)$ in complex molecular systems.

\subsection{Separable theories}
In standard RRKM theory, the dynamics along the reaction coordinate are considered to be separable from those of the other degrees of freedom.
In addition, these dynamics are assumed to be classical, thereby effectively neglecting quantum tunneling and leading to a simple expression for the cumulative reaction probability: \cite{RRKM1,KineticsBook}
\begin{align} \label{NRRKM}
    N_\text{RRKM}^\text{vib}(E)=\sigma\sum_\mathbf{n} P_\text{cl}(E-\mathcal{E}_\text{TS}^{(\mathbf{n})}) ,
\end{align} 
where the classical transmission probability is written in terms of the Heaviside step function as $P_\text{cl}(E) = h(E-V_\text{TS})$, the potential energy of the transition state is $V_{\text{TS}}$ and its vibrational energy levels are
$\mathcal{E}_\text{TS}^{(\textbf{n})}$.
For a molecule with $f$ internal degrees of freedom, there are $f-1$ vibrational modes orthogonal to the reaction coordinate, for which one typically employs the harmonic-oscillator approximation, 
\begin{equation}
\mathcal{E}_\text{TS}^{(\mathbf{n})} = \sum_{j=1}^{f-1} (n_j + \tfrac{1}{2}) \hbar\omega_j^\ddag ,
\end{equation}
where $\omega_j^\ddag$ are the vibrational frequencies at the transition state.
The sum over $\mathbf{n}=\{n_1,\dots,n_{f-1}\}$ in Eq.~\eqref{NRRKM} is understood to be over all vibrational states of the TS such that
$\sum_\mathbf{n} = \sum_{n_{1}=0}^{\infty} \sum_{n_{2}=0}^{\infty} \dots \sum_{n_{f-1}=0}^{\infty}$.
For large polyatomic systems, the number of terms required to converge the sum can become unmanageable such that approximate formulas are often employed to replace the sums by a simpler closed-form expression. \cite{Rabinovitch_DoS_1964}
Such simplifications are, however, only applicable because of the separable approximation made here.
Finally, $\sigma$ is the symmetry factor, which accounts for the number of equivalent reaction mechanisms and is easily determined from the point groups of the reactant and TS geometries. \cite{Turhlar2007Symmetry}

So far we have treated only the vibrational degrees of freedom.
To additionally account for rotational states within the approximation that they are also uncoupled to the reaction coordinate, we can use the $J$-shifting method, \cite{Bowman1991reduced}
\begin{equation}
N_\text{RRKM}(E) = \sum_{J=0}^\infty \sum_{K=-J}^J (2J+1)N_\text{RRKM}^\text{vib}(E - E_{J,K}^\text{rot}),
\label{Jshift}
\end{equation}
where $E_{J,K}^\text{rot}$ are the rotational states of the TS treated as a rigid rotor (in this case assumed to be non-linear and thus labeled by quantum numbers $J$ and $K$), and the factor of $(2J+1)$ accounts for the degeneracy of the molecular orientation. 

The simplest extension of RRKM theory retains the \textit{a priori} separable approximation but includes tunneling effects by employing a one-dimensional quantum transmission probability $P_\text{1D}(E)$ instead of the classical step function $P_\text{cl}(E)$. \cite{Miller1979unimolecular}
The vibrational part of the cumulative reaction probability can then be written in the simple form
\begin{align} 
\label{Nsep}
    N_\text{sep}^\text{vib}(E) = \sigma\sum_\textbf{n} P_\text{1D}(E - \mathcal{E}_\text{TS}^{(\textbf{n})}).
\end{align}
Rotations can again be included here with the $J$-shifting approach as before [Eq.~\eqref{Jshift}].
There are a number of different ways one could choose to define $P_\text{1D}(E)$, such as using analytic expressions for certain barrier shapes.
For example, if the barrier is approximated by an inverted parabola the one-dimensional transmission probability is $P_\text{pb}(E) = [1 + \text{exp}(W_\text{pb}(E)/\hbar)]^{-1}$, where $W_{\text{pb}}(E)={2\pi(V_{\text{TS}}-E)}/{\omega_{\text{TS},0}}$ and $\omega_{\text{TS},0}$ is the magnitude of the imaginary frequency at the TS. \cite{BellBook}
As the perpendicular modes are treated as harmonic oscillators, we call this approach the ``parabolic + harmonic'' (pb + h) approximation.
The related ``Eckart + h" approximation uses the quantum-mechanical expression for transmission through the Eckart barrier \cite{Eckart} fitted to reproduce the barrier height and frequency of the TS. \cite{Johnston1962Eckart,Miller1979unimolecular}

If the barrier shape does not fit these simple analytic forms, it is common to employ a semiclassical (SC) approximation for the reaction probability, $P_\text{1D}(E)$.
The simplest Wenztel--Kramers--Brillouin (WKB) approximation is given by \cite{BellBook}
\begin{equation}
P_{\text{SC}0}(E) =\begin{cases}
               \mathrm{e}^{-W(E)/\hbar},~~~E<V_{\text{TS}}\\
               1,~~~E \ge V_{\text{TS}} ,
            \end{cases}
            \label{PSC0}
\end{equation}
where $W(E)=2\int \sqrt{2m[V(x)-E]}~\text{d}x$ is the abbreviated action along a given tunneling path with potential $V(x)$.
Note that this approximate form has its limitations, most noticeably, $P_{\text{SC0}}(E)$ is inaccurate when $E$ is close to the barrier top. 
To address the former issue, an improved definition combines the semiclassical approximation with the parabolic-top correction, \cite{Faraday,BellBook}
\begin{equation}
P_{\text{SC}}(E) = \begin{cases}
               [1+\mathrm{e}^{W(E)/\hbar}]^{-1},~~~E<V_{\text{TS}}\\
               [1+\mathrm{e}^{W_{\text{pb}}(E)/\hbar}]^{-1},~~~E \ge V_{\text{TS}} .
            \end{cases}
            \label{PSC}
\end{equation}
In many cases, one calculates this tunneling factor along the minimum-energy pathway (MEP), which leads to an approach known as the ``MEP + h" approximation or the zero-curvature tunneling method. \cite{Truhlar_ZCT_1971}
Moreover, there are a number of widely known rate theories \cite{VTSTtunnel} which are based on the semiclassical approximation and perturbative corrections to the separable approximation, such as the small-curvature tunneling method \cite{Truhlar_SCT_1981} and the 1D semiclassical TST. \cite{Greene20161D} 

It is known, however, that in some cases involving deep tunneling or particularly complicated reaction paths, the rate cannot be correctly described within a separable approximation, \cite{Miller1974QTST}
and methods such as these may thus give spurious results.
It has been proposed that the semiclassical approximation to $N(E)$ can instead be computed from instanton theory
without invoking the separable approximation.
\cite{Miller1975semiclassical,Chapman1975rates,Faraday}
We will therefore focus on the instanton approach in this work.

\subsection{Thermal instanton theory}
Within the ring-polymer instanton formalism, \cite{Andersson2009Hmethane,RPInst,Rommel2011locating,Perspective,InstReview}
instanton theory has become a well-established approach for predicting thermal reaction rates with quantum tunneling beyond the separable approximation.
\cite{dimersurf,hexamerprism,i-wat2,porphycene,HCH4,GPR,*Muonium,DaC,Rommel2012enzyme,Meisner2017water,Asgeirsson2018instanton}
There are a number of different formulations of thermal instanton theory which can all be shown to be equivalent in principle. \cite{Miller1975semiclassical,Uses_of_Instantons,Althorpe2011ImF,AdiabaticGreens,InstReview}
We will present the formulation of Miller's original approach, \cite{Miller1975semiclassical} which is most easily connected to the microcanonical theory.

Semiclassical instanton theory is based on an optimal tunneling pathway called the instanton,
which is defined as the cyclic path $\textbf{x}(\tau')$ which makes the Euclidean action
\begin{align}
S\equiv S(\tau)=\int_0^\tau \left[ \tfrac{1}{2} m|\dot{\textbf{x}}(\tau')|^2 + V(\textbf{x}(\tau')) \right] \mathrm{d}\tau'
\end{align}
stationary, where $V(\textbf{x})$ is the potential-energy surface (PES) and $\tau$ the total imaginary time of the periodic orbit.
In this way, the instanton can be characterized by the value of $\tau$.
Alternatively, one can use the instanton energy,
\begin{equation}
E_{\text{I}} = \frac{\partial S}{\partial\tau},
\label{E_inst}
\end{equation}
to characterize the instanton instead of $\tau$,
as there is typically a monotonic relationship between instanton energy and imaginary time (such that larger $\tau$ values correspond to smaller $E_\text{I}$).
In the following, we will make use of this, as it is necessary in order to make a connection to the microcanonical rate.

Another important quantity is related to the fluctuations around the tunneling path, which for a given instanton energy can be characterized by $\omega_j(E_\text{I})=u_j(E_\text{I})/\tau$, where $j=0,1,\dots,f-1$, and $u_j(E_\text{I})$ are the stability parameters of the instanton. \cite{Miller1975semiclassical}
Note that $u_0(E_\text{I})=0$ due to the invariance of the periodic path to cyclic permutations. \cite{Miller1975semiclassical}
Translational and rotational degrees of freedom also lead to zero stability parameters and must thus be treated separately from the vibrations. 
One can define an effective vibrational partition function for the instanton characterized by energy $E_\text{I}$, 
\begin{equation}
Z_{\text{I}}^{\text{vib}}(\beta;E_{\text{I}})= \prod_{j=1}^{f-1}\frac{1}{2\sinh(\beta\hbar\omega_j(E_{\text{I}})/2)}.
\label{ZIvib}
\end{equation}
This expression is written in a general way to allow $\beta$ and $E_\text{I}$ to be chosen independently of each other.
This is a useful extension which we will make use of for our definition of the microcanonical rate, although it is not necessary for the thermal instanton theory presented in this subsection.
For a molecule with rotational degrees of freedom we can define the total rovibrational partition function as
$Z_\text{I}(\beta;E_\text{I}) = Z_{\text{I}}^{\text{vib}}(\beta;E_\text{I}) Z_{\text{I}}^{\text{rot}}(\beta;E_\text{I})$, where
$Z_{\text{I}}^{\text{rot}}(\beta;E_\text{I})$ is defined similarly to a classical rotational partition function for the whole ring-polymer instanton as described in \Ref{InstReview}.

Instantons exist for energies up to $E_\text{I}=V_\text{TS}$, at which point the instanton collapses onto the TS, and the frequencies $\omega_j(E_\text{I})$ reduce to the vibrational frequencies at the TS, $\omega_{\text{TS},j}$ ($j=1,\dots,f-1$).
Thus, these expressions are only defined for $E_{\text{I}}\le V_{\text{TS}}$, but for the convenience of later discussions we extend the definition to include energies above the barrier with $Z_{\text{I}}(\beta;E_\text{I})=Z_{\text{I}}(\beta;V_{\text{TS}})$ for $E_\text{I} > V_\text{TS}$. 

The rate obtained from conventional instanton theory, 
which we refer to as the steepest-descent instanton (SDI) method,
can be derived by taking a number of steepest-descent approximations to the exact thermal rate expression. \cite{Miller1975semiclassical,AdiabaticGreens,InstReview}
At temperature $T$ the rate constant can be calculated from the instanton with period $\tau^*\equiv\beta\hbar$, where  $\beta\equiv\beta(T)$,
and energy $E_\text{I}^*=S'(\tau^*)$ using the following formula: \cite{Miller1975semiclassical,InstReview}
\begin{subequations} \label{kSDI}
\begin{equation}
k_{\text{SDI}}(\beta)= \frac{A_\text{SDI}(\beta)}{Z_{\text{r}}(\beta)} \,\mathrm{e}^{-S(\tau^*)/\hbar}.
\end{equation}
where
\begin{equation}
    A_\text{SDI}(\beta) = \frac{\sigma}{\sqrt{2\pi\hbar}} \left(\frac{\text{d}^2W}{\text{d}E_\text{I}^2}\right)_{E_\text{I}=E_\text{I}^*}^{-\frac{1}{2}} Z_{\text{I}}(\beta;E_\text{I}^*)
\end{equation}
\end{subequations}
Because the period of the instanton is fixed by the temperature, there is a one-to-one correspondence between $\beta$ and $E_\text{I}^*$.
The abbreviated action can therefore be defined by the Legendre transform $W(E_\text{I}^*)=S(\tau^*)-\tau^* E_\text{I}^*$, whose derivatives
 are $W'(E_\text{I}^*) = -\tau^*$ and $W''(E_\text{I}^*) = - S''(\tau^*)^{-1}$.\cite{Kleinert}
The expression in Eq.~\eqref{kSDI} is formally equivalent to the ring-polymer instanton method. \cite{Althorpe2011ImF,AdiabaticGreens,InstReview}

Instantons only exist below a crossover temperature, which is typically estimated by $T_\mathrm{c} \approx \frac{\hbar |\omega_{\text{TS},0}|}{2 \pi k_\text{B}}$, \cite{Benderskii} and thus SDI theory is only valid in this regime.
Above the crossover temperature, classical rate theories are typically reasonably accurate and thus $T_\mathrm{c}$ 
can serve as a good indication of when quantum tunneling becomes important to the thermal rate.
There are a number of scenarios where the SDI rate could predict unphysical or inaccurate results,
such as for tunneling through broad-topped potentials, \cite{broadtop} deep-tunneling in systems with pre-reactive complexes (see Sec.~4.3), or for cases close to the crossover temperature. \cite{Haenggi1988crossover,Zhang2014interpolation,Kryvohuz2011rate,McConnell2017instanton}
We suggest that a good alternative to the SDI method could be based on a microcanonical instanton approximation to $N(E)$, from which the thermal rate could then be simply obtained by evaluating the integral in Eq.~\eqref{kint} numerically.
We call this the thermalized microcanonical instanton (TMI) approach, and it has been shown that similar approaches can overcome some of the shortcomings of the SDI rate. \cite{Faraday,broadtop,McConnell2017microcanonical}

\subsection{Microcanonical instanton theories}
We now consider how the instanton approach can be employed to directly calculate the cumulative reaction probability, $N(E)$.
In contrast to thermal instanton rate theory, applications of instanton theory to microcanonical reactions have been much more limited. \cite{Chapman1975rates,McConnell2017microcanonical,Faraday,Kastner_Methanol_in_Space_2019}
First we describe methods previously proposed and discuss their advantages and disadvantages before introducing our new approach in Sec.~2.4.
Miller formulated the first semiclassical microcanonical theory starting from the Weyl correspondence rule \cite{Miller1975semiclassical}
and a similar theory was more recently re-derived directly as a steepest-descent approximation of the first-principles rate. \cite{AdiabaticGreens}
However, in \Ref{Chapman1975rates} it was argued that this original approach gives poor results for microcanonical rates.
This can be understood by the fact that the expression neglects the ZPE of modes perpendicular to the instanton pathway.
It is possible to correct this major deficiency of the original semiclassical approach by generalizing the first-principles derivation of instanton theory \cite{AdiabaticGreens} with the additional step of treating the fluctuation factors as an effective action when carrying out the steepest-descent integration. \cite{Faraday}
This leads to an approach which is subtly different from Miller's original formulation.  Instead of defining the total energy by the bare instanton energy $E_\text{I}$, the improved approach effectively uses $E=E_{\text{I}}+\mathcal{\bar{E}}_{\text{I}}(\bar{\beta};E_\text{I})$, where the contribution from the perpendicular modes is
\begin{subequations}
\begin{align}
\mathcal{\bar{E}}_{\text{I}} \equiv \mathcal{\bar{E}}_{\text{I}}(\bar{\beta};E_\text{I})=\sum_{j=1}^{f-1}\frac{\hbar\omega_j(E_\text{I})}{2}\coth\left(\frac{\bar{\beta}\hbar\omega_j(E_\text{I})}{2}\right),
\end{align}
and $\bar{\beta}\equiv\bar{\beta}(E_\text{I})=\tau(E_{\text{I}})/\hbar$ is related to the temperature which would correspond to this instanton in a thermal rate calculation.
We have further adapted and simplified this improved semiclassical method from \Ref{Faraday} to give,%
\footnote{The original derivation splits the instanton into two parts, with imaginary time $\tau^+$ and $\tau^-$ which are in general not equal. 
Here, we choose the symmetric split $\tau^+=\tau^-=\tau/2$.
We have also dropped a rather complicated term (which is usually approximately equal to 1) in the prefactor.
}
\begin{align}
N_{\text{OtO}}(E_{\text{I}}+\mathcal{\bar{E}}_{\text{I}})=\sigma Z_{\text{I}}(\bar{\beta};E_{\text{I}}) \mathrm{e}^{+\bar{\beta}\mathcal{\bar{E}}_{\text{I}}} P_{\text{SC}}(E_{\text{I}}),~~~E_{\text{I}}<V_{\text{TS}},
\end{align}%
\label{NIMF}%
\end{subequations}
Note that, in the long imaginary-time limit (or the high-frequency limit), the energy shift becomes equal to the ZPE of the modes perpendicular to the instanton, which we define as $\mathcal{E}_{\text{I}}\equiv\mathcal{E}_{\text{I}}(E_\text{I}) =
\lim_{\bar{\beta}\rightarrow\infty}\mathcal{\bar{E}}_{\text{I}}(\bar{\beta};E_\text{I}) = \sum_{j=1}^{f-1}\tfrac{1}{2}\hbar\omega_j(E_\text{I})$
and the expression reduces to give the dominant first term of the sum in Eq.~\eqref{Nsep}. 
In general however, each mode contributes an effective thermal energy corresponding to the imaginary time of the instanton path.
For comparison, the dominant (single orbit) term of Eq.~(2.39) from \Ref{Miller1975semiclassical} is equivalent to Eq.~(\ref{NIMF}b) with the effective vibrational energy $\bar{\mathcal{E}}_\text{I}=0$.

Like Miller's original formula, each point on the $N(E)$ curve can be evaluated by just one instanton.
We therefore refer to this type of method as a ``one-to-one'' (OtO) approach.
Although OtO methods constitute the most simple microcanonical instanton approaches, they also have some drawbacks, notably that $N_{\text{OtO}}(E)$ is only available for energies where instantons are not collapsed, and it does not connect smoothly to the results of a method valid at higher energies, such as the ``pb + h'' approximation. 
The definition of $N_{\text{OtO}}(E)$ also does not rigorously obey the ``separability principle", i.e.\ $N(E)$ should formally reduce to Eq.~\eqref{Nsep} if the vibrational modes are uncoupled to the reaction coordinate along the tunneling pathway. \cite{Chapman1975rates,Faraday}
Nevertheless, we will show in this work that the approximation is reasonably good for the systems tested.
Miller and coworkers, however, proposed an alternative semiclassical approximation for $N(E)$ which improved upon the original formula and rigorously obeys the separability principle, as well as solving a number of other issues such as connecting smoothly to the classical high-energy regime.
It was not derived directly but rather obtained by taking the pragmatic step of ``unexpanding'' the first two terms of a Taylor series expansion 
\cite{Chapman1975rates} of the original formula in \Ref{Miller1975semiclassical}.
The final expression can be justified in that it has 
similar properties to the separable formula Eq.~\eqref{Nsep} but without explicitly making the separable approximation. 
This alternative approach is no longer a OtO method, but instead involves a direct summation of states over all the excitations of the modes perpendicular to the instantons:
\begin{subequations}
\begin{align}{\label{NE_M_RC}}
N_\text{SC-sum}^\text{vib}(E) &= \sigma\sum_\mathbf{n} P_\text{SC}(E_\text{I}^{(\bf{n})}),
\end{align}
where for each term in the sum, the instanton energy associated with the vibrational state \textbf{n} obeys the transcendental equation
\begin{align}{\label{E_n}}
E_\text{I}^{(\mathbf{n})} = E - \sum_{j=1}^{f-1} (n_j + \tfrac{1}{2})\hbar\omega_j(E_\text{I}^{(\mathbf{n})}).
\end{align}%
\label{directsum}%
\end{subequations}
This approach clearly treats separable systems correctly as then $\omega_j$ becomes independent of energy and $N_\text{SC-sum}^\text{vib}(E)$ reduces to Eq.~\eqref{Nsep}. 
To evaluate $N_\text{SC-sum}^\text{vib}(E)$ for a specific energy, one would need to solve the transcendental equation above for each vibrational configuration $\mathbf{n}$, which is not a simple task as a new instanton must be optimized for every evaluation of $\omega_j(E_\text{I}^{(\mathbf{n})})$.
A more computationally efficient scheme can be defined by first optimizing a set of instantons and using spline interpolation to obtain the action and stability parameters as functions of $E_\text{I}$.
Each configuration $\mathbf{n}$ can then be treated separately by defining a corresponding total energy, $E$, for each instanton via Eq.~\eqref{E_n}.
One then sums over $\mathbf{n}$ up to a maximum excitation number (which should be increased until convergence of the required energy range) to obtain $N_\text{SC-sum}^\text{vib}(E)$.
This approach is feasible for small systems, \cite{McConnell2017microcanonical}
but it can be difficult to treat larger molecules as the computational cost of evaluating the sum over states increases exponentially with the number of vibrational modes
and it is especially difficult to converge when low-frequency (soft) modes are present.
Unfortunately, the terms in the sum here are significantly more complicated than the equivalent terms in RRKM theory, and simple approximations for the sum \cite{Rabinovitch_DoS_1964,BeyerSwineheart} are not applicable to the microcanonical instanton expression.
Rotational degrees of freedom can be treated by the $J$-shifting method [Eq.~\eqref{Jshift}], using the rigid-rotor approximation at the classical TS geometry.
Note that this is a simple approximation to the method proposed in \Ref{Miller1975semiclassical}, which was defined in terms of a sum over the vibrational and rotational quantum numbers simultaneously.  Due to the high density of rotational states, this approximation can drastically reduce the computational cost. 

Finally, we consider a third approach to obtain an instanton approximation to $N(E)$
based on the inverse Laplace transform of the SDI thermal rate [Eq.~\eqref{kSDI}], due to the relation in Eq.~\eqref{kint}.
This is similar to the idea developed by Tao et al.\ for transforming results of ring-polymer molecular dynamics into the microcanonical ensemble \cite{Miller}. 
Of course it is well known that inverse Laplace transforms are not trivial to perform numerically, although a number of approximate approaches 
were explored in \Ref{Miller} which appear to be well behaved.
The method we choose to use is the stationary-phase approximation to the inverse Laplace transform (SPA-ILT),
partly because of its simplicity, but also because it seems consistent with the steepest-descent approximations already inherent in instanton theory.
Exploring other numerical methods for implementing the ILT, which may lead to increased accuracy, will be left for future work. 
We discuss SPA-ILT in detail in Appendix A and applying this approach to Eq.~\eqref{kint}, we obtain
\begin{subequations}
\begin{equation}
N_{\text{ILT-SDI}}(E) =\sqrt{2\pi\hbar^2} \left(\frac{\partial^2\ln [A_\text{SDI}(\beta)\mathrm{e}^{-S(\beta\hbar)/\hbar}]}{\partial\beta^2}\right)_{\beta=\beta_{\text{sp}}}^{-\frac{1}{2}}A_\text{SDI}(\beta_\text{sp})\mathrm{e}^{-S(\beta_\text{sp}\hbar)/\hbar} \mathrm{e}^{\beta_{\text{sp}}E} ,
\label{ILT_SDI}
\end{equation}
where
\begin{equation}
E \equiv E(\beta_{\text{sp}})=-\left(\frac{\partial\ln [A_\text{SDI}(\beta)\mathrm{e}^{-S(\beta\hbar)/\hbar}]}{\partial\beta}\right)_{\beta=\beta_{\text{sp}}},~~{\beta_{\text{sp}}}\in\mathbb{R}^+
\label{ILT_SDI2}
\end{equation}
\end{subequations}
In these expressions, $\beta_{\text{sp}}$ is a free variable which defines a corresponding total energy, $E$, via Eq.~\eqref{ILT_SDI2}.
The derivatives with respect to $\beta$ can be performed by fitting a spline interpolation to a set of SDI calculations at different temperatures.
One can calculate $N_{\text{ILT-SDI}}(E)$ over a desired range of energies by using the corresponding range of $\beta_\text{sp}$ values. 
Using the relation in Eq.~\eqref{E_inst}, one finds that the total energy is approximately $E\approx E_\text{I}^*+\mathcal{\bar{E}}_\text{I}(\beta_\text{sp};E_\text{I}^*)+f^\text{rot}/2\beta_\text{sp}$ (dropping a small term related to the second derivative of $W$), where $E_\text{I}^*$ is the energy of the instanton with $\tau^*=\beta_\text{sp}\hbar$, and $f^\text{rot}$ is the number of rotational modes of the system (equal to 2 or 3 depending on whether the ring-polymer instanton is linear or nonlinear) which are treated classically and hence obey the equipartition theorem. 
The total energy thus takes account of the vibrational and rotational excitations of the instanton, at least in an average manner.
Similar to Eq.~\eqref{NIMF}, each instanton calculation returns one point on the $N_{\text{ILT-SDI}}(E)$ curve, and this is thus another example of an OtO method.
%
It provides a simple way to convert thermal SDI rate directly into a microcanonical instanton rate and is thus only reliable in cases where SDI itself is valid. 
For example, as shown in \Ref{Faraday}, SDI can be derived from $P_\text{SC0}$ rather than $P_\text{SC}$, meaning that this method will usually be less accurate for
studying shallow tunneling near the barrier top.
Also since SDI rates only exist below the crossover temperature, $N_{\text{ILT-SDI}}(E)$ is not available at all for high energies, and may break down for broad-top barriers. \cite{broadtop}

In summary, the three methods introduced in this section each have different advantages and disadvantages.
Miller's direct-sum method has generally been assumed to be the most reliable, as it is the only one which obeys the separability principle and tends correctly to a reasonable classical result in the high-energy regime.
It is, however, unfortunately more computationally expensive to evaluate than the other two.
In this work we will compare these three microcanonical instanton approaches, and generally find that they are in good agreement with each other for the unimolecular and bimolecular gas-phase reactions tested, at least within their ranges of applicability.

\subsection{Density-of-states instanton theory}
The main theoretical development described in this work is a new implementation of microcanonical instanton theory
related to Miller's approach, \cite{Chapman1975rates} but which avoids the intimidating sum over all the vibrational and rotational states.
In this new method, we assign a density of states (DoS) to each instanton pathway
and are then able to 
replace the direct sum with an integral over instantons: 
\begin{subequations}
\begin{equation}
N_\text{DoS}(E)=\sigma\int_{E_\text{I}^\text{min}}^{\infty}\text{d}E_{\text{I}}~P_{\text{SC}}(E_{\text{I}})\rho_\text{I}(E;E_{\text{I}}),
\label{TMI1}
\end{equation} 
where $E_\text{I}^\text{min}$ is the lowest energy of an instanton that exists in the system.  $\rho_\text{I}(E;E_{\text{I}})$ is the DoS of the vibrational and rotational excitations of the modes perpendicular to the instanton with energy $E_\text{I}$, defined via the Laplace transform:
\begin{equation}
\int_{E_\text{I}}^{\infty}\text{d}E~\rho_\text{I}(E;E_{\text{I}}) \mathrm{e}^{-\beta E} = Z_{\text{I}}(\beta;E_{\text{I}})\mathrm{e}^{-\beta E_{\text{I}}}.
\label{TMI2}
\end{equation}%
\label{TMI}%
\end{subequations}
This is a commonly used trick to avoid summation over states in quantum mechanics, and it retains the desirable properties of Miller's direct summation approach (i.e.\ it obeys the separable principle and gives the correct high-energy classical limit).
The trade off, however, is that the DoS is not trivial to compute exactly.
We therefore again employ the stationary-phase approximation (see Appendix A), partly because it is consistent with the steepest-descent approximations inherent to the instanton theory, and compute $\rho_\text{I}(E;E_{\text{I}})$ as
\begin{subequations}
\begin{equation}
\rho_\text{I}(E;E_{\text{I}}) = \left(2\pi\frac{\partial^2\ln [Z_{\text{I}}(\beta;E_{\text{I}})\mathrm{e}^{-\beta E_{\text{I}}}]}{\partial\beta^2}\right)_{\beta=\beta_{\text{sp}}}^{-\frac{1}{2}}\mathrm{e}^{\beta_{\text{sp}}E}Z_{\text{I}}(\beta_{\text{sp}};E_{\text{I}})\mathrm{e}^{-\beta_{\text{sp}}E_{\text{I}}},
\end{equation}
where the total energy, $E$, is a function of the free variable $\beta_{\text{sp}}$,
\begin{equation}
\begin{split}
E\equiv E(\beta_\text{sp};E_\text{I}) &= -\left(\frac{\partial\ln [Z_{\text{I}}(\beta;E_{\text{I}})\mathrm{e}^{-\beta E_{\text{I}}}]}{\partial\beta}\right)_{\beta=\beta_{\text{sp}}}\\
&=E_\text{I}+\mathcal{\bar{E}}_\text{I}(\beta_\text{sp};E_\text{I})+\frac{f^\text{rot}}{2\beta_\text{sp}},~~{\beta_{\text{sp}}}\in\mathbb{R}^+.
\end{split}
\end{equation}
\label{ILTsp}
\end{subequations}
Note that $\beta_\text{sp}$ is not related to the imaginary time of the instanton periodic orbits.
By definition, $\rho_\text{I}(E,E_\text{I})=0$ for energies below the ZPE-corrected instanton energy (i.e.\ for $E<E_\text{I}+\mathcal{E}_\text{I}$) as here the density of states corresponding to the instanton is zero.
Implementation of the DoS instanton method requires nothing more than simple, computationally inexpensive post-processing of the data obtained from a handful of ring-polymer instanton calculations with different imaginary times, $\tau$.
We briefly outline the procedure here and give further technical details in Sec.~3\@.
For each instanton optimized at a given value of $\tau$, one can compute $E_\text{I}$ and $\omega_j(E_\text{I})$.  Then, given a set of values of the free variable $\beta_{\text{sp}}$ one can evaluate the corresponding energies $E\equiv E(\beta_\text{sp};E_\text{I})$ using Eq.~(\ref{ILTsp}b).
Then Eq.~(\ref{ILTsp}a) allows one to compute the density of states, $\rho_\text{I}(E;E_{\text{I}})$, over this set of energies, $E$, for an instanton characterized by $E_\text{I}$.
Finally, $N_\text{DoS}(E)$ can be obtained by numerical integration over the instantons characterized by different $E_\text{I}$ values using Eq.~\eqref{TMI1}.
Once $N_\text{DoS}(E)$ is obtained, the microcanonical rate can be computed from Eq.~\eqref{kMI} and the thermal rate from Eq.~\eqref{kint} with integrals carried out numerically using Simpson's rule. 
It is worth noting that the $N(E)$ for the separable approximations (pb + h, Eckart + h, MEP + h, etc.) can be computed in a similar fashion as in Eqs.~\eqref{TMI} and \eqref{ILTsp}.
To do this, one can simply replace $P_\text{SC}(E_\text{I})$ with the corresponding $P_\text{1D}(E_\text{I})$ and $Z_\text{I}(\beta;E_\text{I})$ with $Z_\text{TS}(\beta)$ in Eqs.~\eqref{TMI} and \eqref{ILTsp}.
This can be viewed as replacing the sum in Eq.~\eqref{Nsep} with an integration over the density of states obtained from the stationary-phase approximation.
We have checked that this approach is in close agreement with the direct-sum method, which confirms that the stationary-phase approximations involved in the new procedure are valid for typical systems.

In the high energy limit, $P_\text{SC}(E_\text{I})$ becomes the transmission probability of a parabolic barrier, and $Z_\text{I}(\beta;E_\text{I})$ becomes $Z_\text{TS}(\beta)$.
Therefore, as discussed above, $N_\text{DoS}(E)$ reduces to the ``pb + h'' approximation at high energies.
Since the ``pb + h'' approximation is typically considered to be good in this regime, this is a desirable feature of the DoS instanton method.

\subsection{Thermalized Microcanonical Instanton Theory}
Here we demonstrate important properties of the DoS instanton method when it is used to obtain a thermal rate through the TMI formalism.
The thermal rate at temperature $T$ under this approximation is given by
\begin{subequations}
\begin{alignat}{3}
k_{\text{TMI}}(\beta)
&=\frac{\sigma}{2\pi\hbar}\frac{1}{Z_{\text{r}}(\beta)}\int_{\mathcal{E}_{\text{r}}}^{\infty}\text{d}E~\mathrm{e}^{-\beta E}\int_{E_\text{I}^\text{min}}^{\infty}\text{d}E_{\text{I}}~P_{\text{SC}}(E_{\text{I}})\rho_\text{I}(E;E_{\text{I}})\\
    & = \frac{\sigma}{2\pi\hbar}\frac{1}{Z_{\text{r}}(\beta)} \int_{E_\text{I}^\text{min}}^{\infty}\text{d}E_{\text{I}}~P_{\text{SC}}(E_{\text{I}}) \int_{\mathcal{E}_{\text{r}}}^{\infty}\text{d}E~\rho_\text{I}(E;E_{\text{I}})\mathrm{e}^{-\beta E}\\
    & = \frac{\sigma}{2\pi\hbar}\frac{1}{Z_{\text{r}}(\beta)} \left[ \int_{\tilde{E_\text{I}}}^{\infty}\text{d}E_{\text{I}}~P_\text{SC}(E_\text{I})Z_{\text{I}}(\beta;E_{\text{I}})\mathrm{e}^{-\beta E_{\text{I}}} + \mathcal{T}(\beta;\tilde{E_\text{I}}) \right],
\end{alignat}
\label{TMI_derive}
\end{subequations}
where $\beta\equiv\beta(T)$.

Equation~(\ref{TMI_derive}c) partitions the TMI rate into two parts.  
The first term describes tunneling via instantons whose entire density of states lies above the reactant ZPE\@.
%
The lower limit is thus defined such that $\tilde{E_\text{I}}$ is the energy of an instanton that satisfies $\tilde{E_\text{I}}+\mathcal{E}_\text{I}(\tilde{E_\text{I}})=\mathcal{E}_\text{r}$.
The second term, 
\begin{equation}
\mathcal{T}(\beta;\tilde{E_{\text{I}}}) = \int_{E_\text{I}^\text{min}}^{\tilde{E_{\text{I}}}}\text{d}E_{\text{I}}~P_\text{SC}(E_\text{I})\int_{\mathcal{E}_{\text{r}}}^{\infty}\text{d}E~\rho_\text{I}(E;E_{\text{I}})\,\mathrm{e}^{-\beta E},
\end{equation}
represents the contribution from ``sub-energetic'' instantons with $E_I < \tilde{E_\text{I}}$.
\footnote{Note that in complex systems with more than one barrier, more than one solution of $\tilde{E_\text{I}}$ might be found, and 
in this case, one should simply integrate over all instantons with $E_\text{I}+\mathcal{E}_\text{I}(E_\text{I})< \mathcal{E}_\text{r}$ instead.}

\begin{figure}[!ht]
    \centering
    \includegraphics[width=8cm]{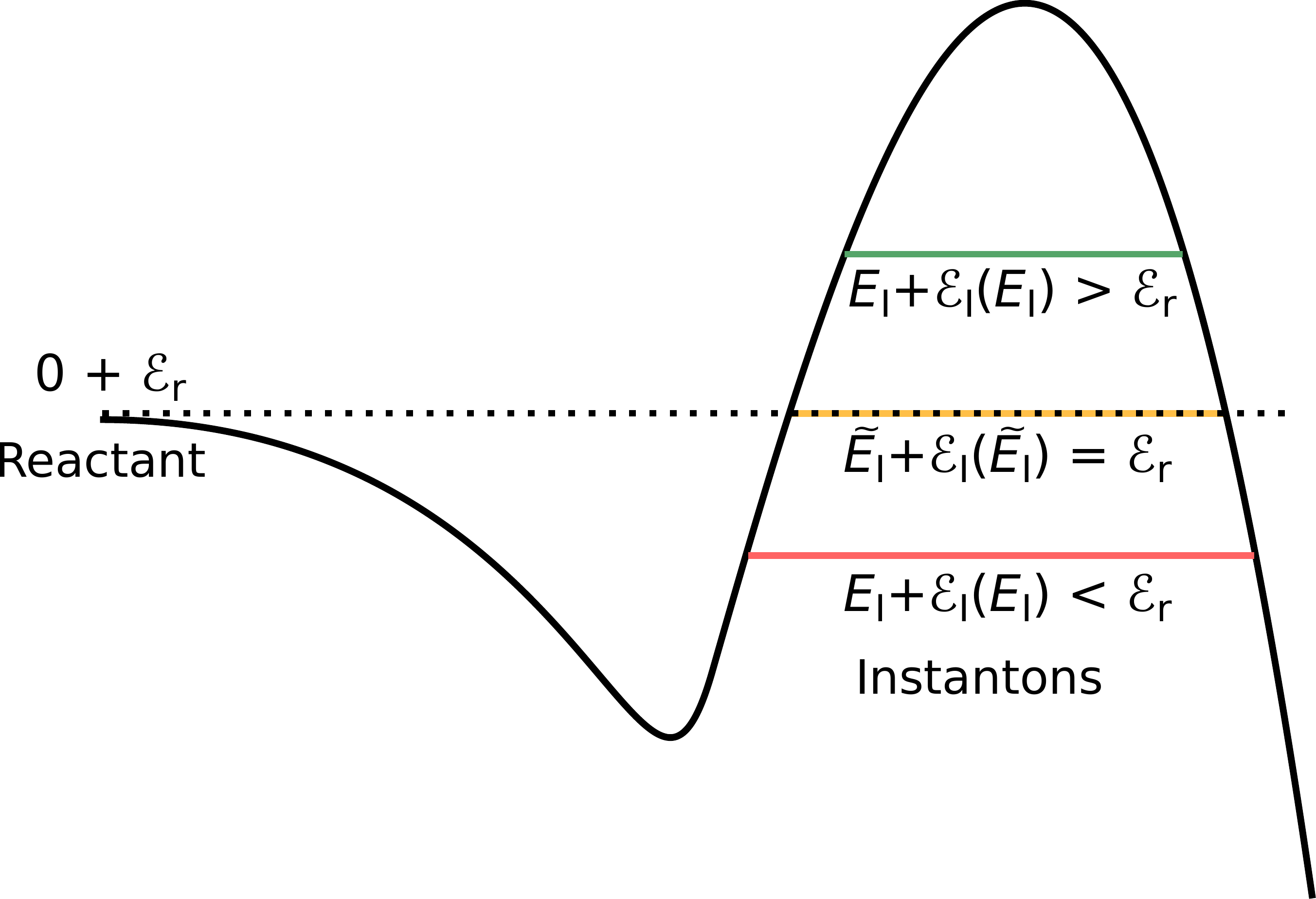}
    \caption{
    Illustration of the normal instanton above the reactant energy (green), sub-energetic instanton below that (red), and the borderline case $E_\text{I}=\tilde{E_\text{I}}$ (orange).
    It is demonstrated here for a bimolecular reaction with a pre-reactive minimum, but note that the three types of instantons may also 
    be present in more general unimolecular and elementary bimolecular reactions if the ZPE at the TS is lower than in the reactant.
    }
    \label{EIstart}
\end{figure}

In many cases, it may not be possible to find a solution for $\tilde{E_\text{I}}$ at all,
and $\tilde{E_\text{I}}$ should be set to $E_\text{I}^\text{min}$ and $\mathcal{T}$ to zero.
However, there are also situations where this is not the case; this is most commonly encountered in reactions with pre-reactive complexes, as depicted in Fig.~\ref{EIstart}.
Within the energy redistribution assumption of the RRKM theory, even the sub-energetic pathways should also contribute to the reaction rate 
but must be treated carefully to ensure that only excitations with total energies above the reactant ZPE are included in the integral.
In these cases, 
we cannot simply integrate over $E$ using Eq.~\eqref{TMI2} as in the first term because
the full range of the instanton's DoS is no longer within the limits of the integral.
The integral can however be evaluated numerically and 
the lower limit of $\mathcal{E}_\text{r}$ ensures
that we only consider reactions which occur at total energies above the reactant ZPE\@.
Typically, the contribution to the thermal rate from this term only becomes important at very low temperature,
when these low-energy tunneling pathways dominate.
We will discuss this term further in Sec.~4.3 when we study a reaction which demonstrates these effects.

To study the classical limit of Eq.~(\ref{TMI_derive}c), we can replace $P_{\text{SC}}(E_\text{I})$ by its classical expression $h(E_\text{I}-V_\text{TS})$ and note that $\mathcal{T}$ is zero in this case.
Equation~\eqref{TMI_derive} then reduces to the Eyring TST rate expression as expected:
\begin{equation}
\begin{split}
k_{\text{TMI,cl}}(\beta) &= \frac{\sigma}{2\pi\hbar}\frac{1}{Z_{\text{r}}(\beta)}\int_{V_\text{TS}}^{\infty}\text{d}E_{\text{I}}~Z_\text{I}(\beta;V_\text{TS})\mathrm{e}^{-\beta E_\text{I}}\\
&=\frac{\sigma}{2\pi\beta\hbar}\frac{Z_\text{TS}(\beta)}{Z_{\text{r}}(\beta)}\mathrm{e}^{-\beta V_\text{TS}} \equiv k_{\text{TST}}(\beta),
\end{split}
\label{kEyring}
\end{equation}
where $Z_\text{TS}(\beta)=Z_\text{I}(\beta;V_\text{TS})$ is the partition function of the TS\@. 
In another important limit, when the modes orthogonal to the reaction coordinate are assumed to be separable along the tunneling pathways, TMI under the new formulation reduces correctly to a one-dimensional expression. 
In this case, the perpendicular vibrational frequencies $\omega_j$ do not change along the reaction coordinate (i.e.\ $\omega_j(E_\text{I})=\omega_{\text{TS},j},~j=1,...,f-1$), which leads to
\begin{equation}
k_{\text{TMI,sep}}(\beta) = \frac{\sigma}{2\pi\hbar}\frac{1}{Z_{\text{r}}(\beta)}\left[Z_{\text{TS}}(\beta)\int_{\tilde{E_\text{I}}}^{\infty}\text{d}E_{\text{I}}~P_\text{SC}(E_\text{I})\,\mathrm{e}^{-\beta E_{\text{I}}}+\mathcal{T}(\beta;\tilde{E_\text{I}})\right] .
\label{TMI_sep}
\end{equation}
One can see that the first term is the standard formula for the TST rate corrected by a one-dimensional semiclassical tunneling factor. \cite{VTSTtunnel} 
The second term only makes a significant contribution when the reaction is dominated by tunneling through the bottom of the barrier
and is valid within the RRKM assumption that the vibrational energy is completely redistributed.
Finally we show that one can also recover the conventional SDI rate as an approximation to the new TMI formula.
We first replace $P_\text{SC}(E_\text{I})$ with $P_\text{SC0}(E_\text{I})$, which is valid for energies well below the barrier top.
We also assume the rate is dominated by energies 
not too close to the bottom of the barrier
and can thus neglect the $\mathcal{T}$ term
to obtain
\begin{equation}
k_{\text{TMI}}(\beta)\approx\frac{\sigma}{2\pi\hbar}\frac{1}{Z_{\text{r}}}\int_{\tilde{E_\text{I}}}^{\infty}\text{d}E_{\text{I}}~Z_{\text{I}}(\beta;E_{\text{I}})\mathrm{e}^{-[W(E_\text{I})-\beta\hbar E_{\text{I}}]/\hbar} .
\label{TMI_SDI1}
\end{equation}
Next, we assume that $Z_{\text{I}}(\beta;E_{\text{I}})$ depends only weakly on $E_\text{I}$ compared to the other terms, which is valid when the instanton vibrational frequencies, $\omega_j$, vary slowly with respect to $E_{\text{I}}$.
Then, taking the steepest-descent approximation for the integration over $E_{\text{I}}$, we obtain
\begin{equation}
k_{\text{TMI}}(\beta) \approx\frac{\sigma}{\sqrt{2\pi\hbar}}\frac{Z_{\text{I}}(\beta;E_\text{I}^*)}{Z_{\text{r}}}\left(\frac{\text{d}^2W}{\text{d}E_\text{I}^2}\right)_{E_\text{I}=E_\text{I}^*}^{-\frac{1}{2}}\mathrm{e}^{-[W(E_\text{I}^*)+\beta\hbar E_\text{I}^*]/\hbar} \equiv k_{\text{SDI}}(\beta).
\end{equation}
Here, $E_\text{I}^*$ is the minimum of $W(E_\text{I}^*)+\beta\hbar E_\text{I}^* = S(\beta\hbar)$, meaning that $W'(E_\text{I}^*) = -\beta\hbar$.
This proof holds as long as $E_I^*$ occurs within the limits of the $E_I$ integral in Eq. \eqref{TMI_SDI1}.
Therefore, TMI reduces to SDI as long as the dominant tunneling takes place neither too close to the top nor the bottom of the barrier.
In summary, the above analysis shows that the new formulation of TMI satisfies the quantum mechanical separation principle for scattering reactions with uncoupled perpendicular modes, that it tends to the correct classical limit, and that it is strongly related to the conventional SDI rate when appropriate.
In general, however, we expect that TMI should outperform SDI and this will be demonstrated in Sec.~4.3.

\section{Computational details}
To apply the new microcanonical DoS instanton method to molecular reactions, we carry out the following steps.
First we optimize a set of instantons with different imaginary times, $\tau$, to obtain a range of $E_\text{I}$. 
The number of instantons required for our method is not at all excessive; for each of the examples, fewer than ten were optimized to converge the final result.
Although specialized methods exist for locating instantons with a specified energy, \cite{Einarsdottir2012path,Asgeirsson2018instanton}
for our work we have found it simplest to optimize the path for a given value of $\tau$ and determine the corresponding energy $E_\text{I}$ afterwards.
Each instanton was thus optimized using standard methods in the discrete path (ring-polymer) representation. \cite{Rommel2011locating,InstReview}
To implement the DoS instanton method, one can see from Eq.~\eqref{TMI2} that $Z_{\text{I}}(\beta;E_{\text{I}})$ needs to be calculated for any $\beta$ over continuous $E_{\text{I}}$ values.
This means that the new method also requires the stability parameters of instanton trajectories [see Eq.~\eqref{ZIvib}], and therefore Hessians along the paths.
We used the conventional method of computing $\omega_j$ for each instanton from the eigenvalues of the Monodromy matrix. \cite{Miller1975semiclassical}
To obtain $\omega_j$ as a continuous function of $E_{\text{I}}$, we used splines to interpolate the stability parameters calculated from individual instantons.
It has been shown that the stability parameters can sometimes be numerically unstable for long imaginary time instanton trajectories. \cite{DaC,Loehle2018stability}
Approximate methods have been proposed to calculate the stability parameters in these cases, such as eigenvalue tracing, which sacrifice accuracy for numerical stability.\cite{McConnell2017microcanonical,Loehle2018stability}
We did not find it necessary to use these approximations for the calculations presented in this paper.
Instead, to obtain the low-energy limit of the stability parameter, $\omega_j(0)$, we simply used the fact that this corresponds to an instanton path with infinite imaginary time, which would have almost all its beads in the reactant state (assuming an exothermic reaction).
This means that $\omega_j(0)^2$ are the eigenvalues of the mass-weighted Hessian of the reactant with the mode along the lowest-energy instanton path projected out. 
Next we compute the DoS of the instantons, $\rho_\text{I}(E;E_{\text{I}})$, using SPA-ILT\@. 
Only interpolated properties of instantons are used for this part of the calculation, and no extra instanton optimizations are required. 
For a given $E_\text{I}$ value, we select a $\beta_{\text{sp}}$ grid with 500 points, and evaluate $E$ on this grid using Eq.~(\ref{ILTsp}b). 
Then we calculate $\rho_\text{I}(E;E_{\text{I}})$ using Eq.~(\ref{ILTsp}b) for all $E$ values obtained, and a smooth $\ln[\rho_\text{I}(E;E_{\text{I}})]$ function is obtained by spline interpolation over the discrete $E$ values. 
The results should be converged with respect to the size of the $\beta_{\text{sp}}$ grid.
As a practical guideline, the $E$ values evaluated from the $\beta_{\text{sp}}$ grid range should start close to $E_{\text{I}}+\mathcal{E}_{\text{I}}$ and run at least as high as values of $E$ at which we intend to compute $N_\text{DoS}(E)$.
The above process is repeated for a grid of $E_{\text{I}}$ values with at least 400 points spanning the range from the lowest $E_{\text{I}}$ of the instantons to the largest $E$ value of the $N_\text{DoS}(E)$ curve we want to obtain.
Again, increasing the $E_{\text{I}}$ grid size does not require instanton optimizations thus does not come with any significant computational cost.
A continuous function of $W(E_{\text{I}})$ can also be obtained by spline interpolation of the abbreviated actions of the instantons.
Finally $N_\text{DoS}(E)$ is obtained via numerical integration over the $E_{\text{I}}$ grid [Eq.~\eqref{TMI1}] using Simpson's rule.
The parabolic-top corrected $P_\text{SC}(E_\text{I})$ (see Eq.~\eqref{PSC}) is used in all the calculations.
We present results in Sec.~4 for three chemical reactions.
For comparison, $N(E)$ is also computed using other methods.
The ``pb + h'' method can also be calculated using SPA-ILT similar to Eq.~\eqref{ILT_SDI}, by replacing factors from $k_\text{SDI}(\beta)$ by the corresponding factors from  $k_{\text{pb}}(\beta)=\frac{\beta\hbar\omega_{\text{TS}}/2}{\sin(\beta\hbar\omega_{\text{TS}}/2)}k_{\text{TST}}(\beta)$. \cite{BellBook}
All the methods within the separable approximation, such as the ``Eckart + h'' and the ``MEP + h'' methods, can be computed directly from Eq.~\eqref{Nsep} using the direct summation approach with $J$-shifting up to a certain cutoff, $J_\text{max}$, on the order of 100--300.
One can also compute them using the DoS formulation and SPA-ILT, which gives almost identical results.
For the ILT-SDI method [Eq.~\eqref{ILT_SDI}], we calculate $\ln[A_\text{SDI}(\beta)\mathrm{e}^{-S(\beta\hbar)/\hbar}]$ from instantons with different $\tau=\beta\hbar$, and then use spline interpolation to obtain a smooth function.
For the smallest system studied, we also computed the instanton $N(E)$ from Eq.~\eqref{NE_M_RC} using the direct summation method.

\section{Results and discussion}
\subsection{H + H$_2$ reaction}
First we test our new approach on a well understood elementary bimolecular reaction, \ce{H + H2 -> H2 + H}, which proceeds directly without any intermediates or pre-reactive complexes.
For this system, the quantum-mechanical $N(E)$ result has been obtained using the log-derivative method \cite{Skouteris2000ABC}
on the LSTH potential-energy surface, \cite{LSTH2}
which we take as our exact benchmark.
Instanton calculations were carried out on the same PES in Cartesian coordinates (9 degrees of freedom)
and eight instantons were optimized, each with 128 ring-polymer beads and corresponding to temperatures between 52 K and 348 K such that they were roughly evenly spaced in energy.
The symmetry factor is $\sigma=2$ for this reaction.

\begin{figure}[!ht]
    \centering
    \includegraphics[width=9cm]{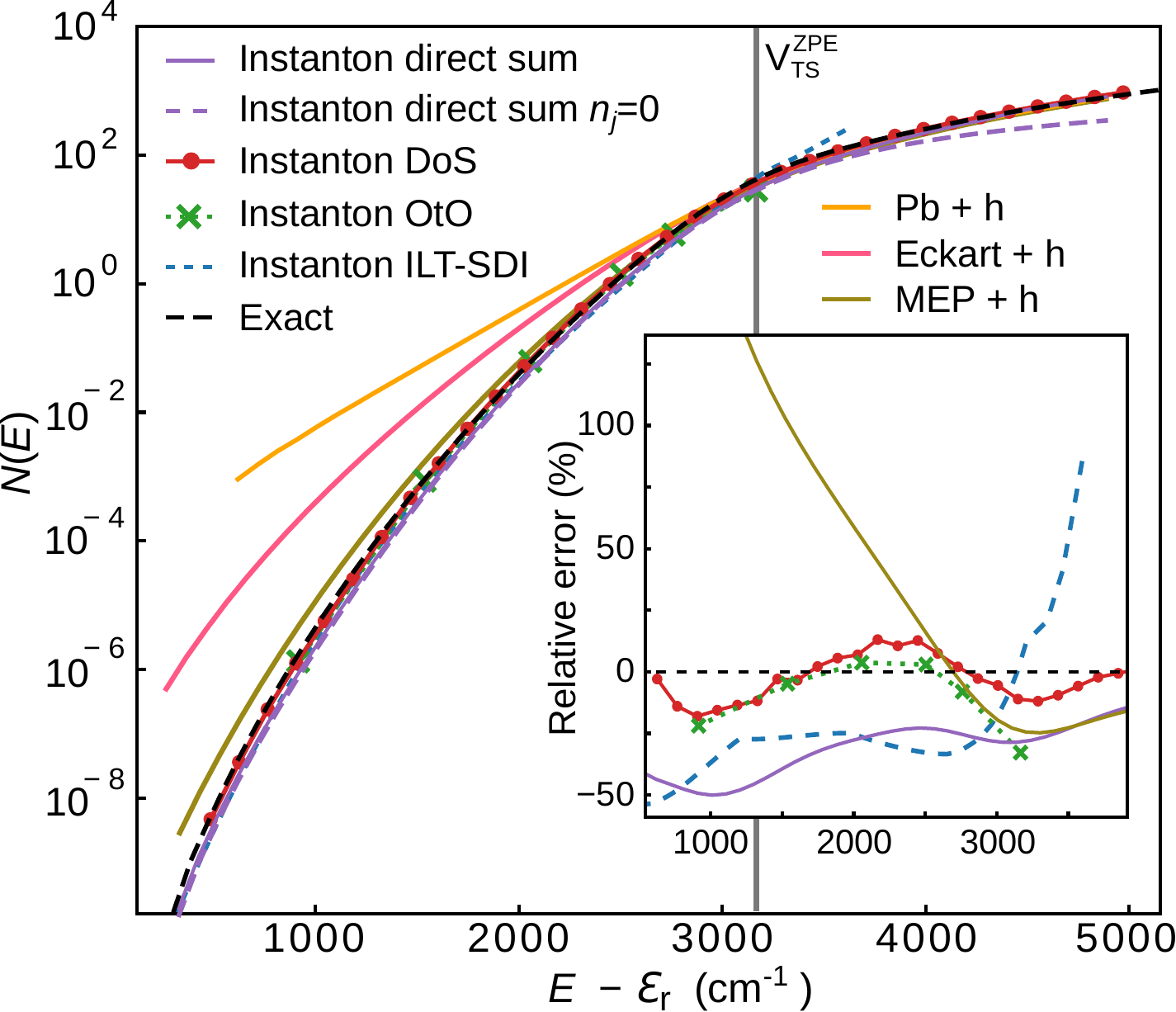}
    \caption{
    $N(E)$ for the full-dimensional H + H$_2$ reaction, calculated with various methods.
    The direct-sum approach used up to 3 excited states for each of the bending modes (low frequency) but just the ground state for the stretching mode (high frequency).
    The dashed purple line shows the result of taking the ground state only for all degrees of freedom.
    The ZPE-corrected TS energy, $V_\text{TS}^\text{ZPE}=3174~\text{cm}^{-1}$, is also indicated.
    The inset shows the percentage error with respect to the exact result.
    }
    \label{HH2_NE}
\end{figure}

A comparison of the results from various methods is shown in Fig.~\ref{HH2_NE}.
It can be seen that in the low-energy tunneling regime, all the instanton-based methods give similar predictions and are in good agreement with the exact result.
The new method, labeled instanton DoS, is not only accurate (within only 20\% error) at describing deep tunneling well below the barrier, but is also in good agreement with the exact result for the shallow tunneling near the barrier top and in the classical regime above the barrier.
It is also in good agreement with the direct-sum method, on which it is based, which demonstrates that the approximate SPA-ILT used is valid.
The small differences between the direct-sum and the DoS methods are due to a combination of the $J$-shifting approximation in the former and the stationary-phase approximation in the latter, neither of which appears to lead to a significant discrepancy.
It is at first surprising that the SPA-ILT works so well for such a small system with only a few vibrational degrees of freedom, since it is expected to be accurate only when the density of states is high.
The numerical SPA-ILT works well however, in this case because of the existence of the quasi-continuous rotational states.

To understand the importance of vibrational excitations to the cumulative reaction probability, we performed a comparison with a calculation based only on the vibrational ground state of the instanton.
This is a brutal truncation of the sum in Eq.~\eqref{directsum} that keeps only the first term,
\begin{equation} 
N_{\text{SC},\mathbf{n}=0}^\text{vib}(E) = \sigma P_\text{SC}(E - \mathcal{E}_\text{I}),
\label{nv0}
\end{equation}
although all rotational states are still included using Eq.~\eqref{Jshift}.
One can see that in this system, for $E$ near and below the ZPE-corrected barrier top, $V_\text{TS}^\text{ZPE}$, including the vibrationally excited states made little to no difference compared to using the vibrational ground state only.
It is only at high energies ($E>4000$ cm$^{-1}$) that the vibrationally excited states start to play a noticeable role.
This behavior is expected in small molecules, but may no longer be true for larger molecules which include low-frequency modes.
We also present results from the two simpler instanton approximations.
The OtO instanton method is in good agreement with the DoS method (except near the barrier top where it is known to break down), even though the two methods were derived from different principles.
The simple ILT-SDI method also works well for this reaction, because it relies on the standard SDI theory, which, as is well known, gives an accurate thermal rate for deep tunneling in this system
\cite{Ceotto2012instanton,InstReview,Kryvohuz2011rate,MUSTreview}.
Nevertheless, this method shows a slightly higher relative error of 30--40\%, and breaks down for energies near or above $V_\text{TS}^\text{ZPE}$ due to well-known problems of SDI near the crossover temperature.

In contrast to the instanton approaches, the methods based on the separable approximation perform less well.
The ``pb + h'' method is reliable only in the classical over-the-barrier regime, but drastically overestimates the $N(E)$ below the barrier.
This is an indication of how strongly the real system deviates from the simple separable model with a parabolic shape. 
If we use the ``Eckart + h'' method instead, the overestimation of $N(E)$ is significantly reduced, but it is still orders of magnitude off from the exact result.
As expected, the Eckart model is a better description of the real barrier than a parabola, but it still models this system poorly.
Finally, using the ``MEP + h'' method further improves the predicted curve, but one can see that it still leads to a less accurate prediction than can be obtained from instanton theory in the deep-tunneling regime and that the error increases as the energy is lowered.
It is interesting that the ``MEP + h'' method overestimates $N(E)$, whereas one might intuitively have expected that it should underestimate it due to its neglect of corner-cutting effects.
Further analysis, as detailed in the supporting information (SI), shows that the overestimation is actually caused by the separable approximation itself, and that the corner-cutting effects for this system are relatively small.
Instanton theory, on the other hand, takes into account non-separable effects, corner-cutting, as well as an accurate description of the barrier shape and is thus the most reliable method overall.

\subsection{Unimolecular reaction of a Criegee Intermediate}
Criegee intermediates, which contain a \ch{R2COO} backbone, play an important role in atmospheric chemistry. \cite{Criegee1975, Criegee_NH3, Criegee_network}
Their reactivity, especially in the form of unimolecular hydrogen transfers, contributes to the formation and cycling of \ch{OH} radicals, which are one of the most important atmospheric oxidants.
Here we study the microcanonical rate for the unimolecular hydrogen transfer between the methyl carbon and the terminal oxygen in a simple Criegee intermediate, \textit{syn-}\ch{CH3CHOO}, and compare our predicted rate constants with those from recent experiments. \cite{Fang2016CH3CHOO, Kidwell2016CH3CHOO}
These experiments used an infrared laser to excite the molecule to a target energy and then measured the time dependence of the signal of the product, from which the microcanonical rate can be obtained.  
Because the excitation was to an energy below the reaction barrier, it is clear that the hydrogen-transfer mechanism proceeds via a tunneling mechanism. 

We employed a PES for this molecule which was fitted by Bowman and coworkers using CCSD(T)-F12b calculations and which is expected to be accurate to $\sim 60 \, \mathrm{cm^{-1}}$. \cite{Kidwell2016CH3CHOO}
A total of seven instantons were optimized, each with 128 ring-polymer beads and with $\tau$ values corresponding to the temperature range 136 K to 380 K such that they were again roughly evenly spaced in energy. \cite{PierreThesis}
Due to the symmetry of the molecule, $\sigma=2$ for this reaction.

\begin{figure}[!ht]
    \centering
    \includegraphics[width=9cm]{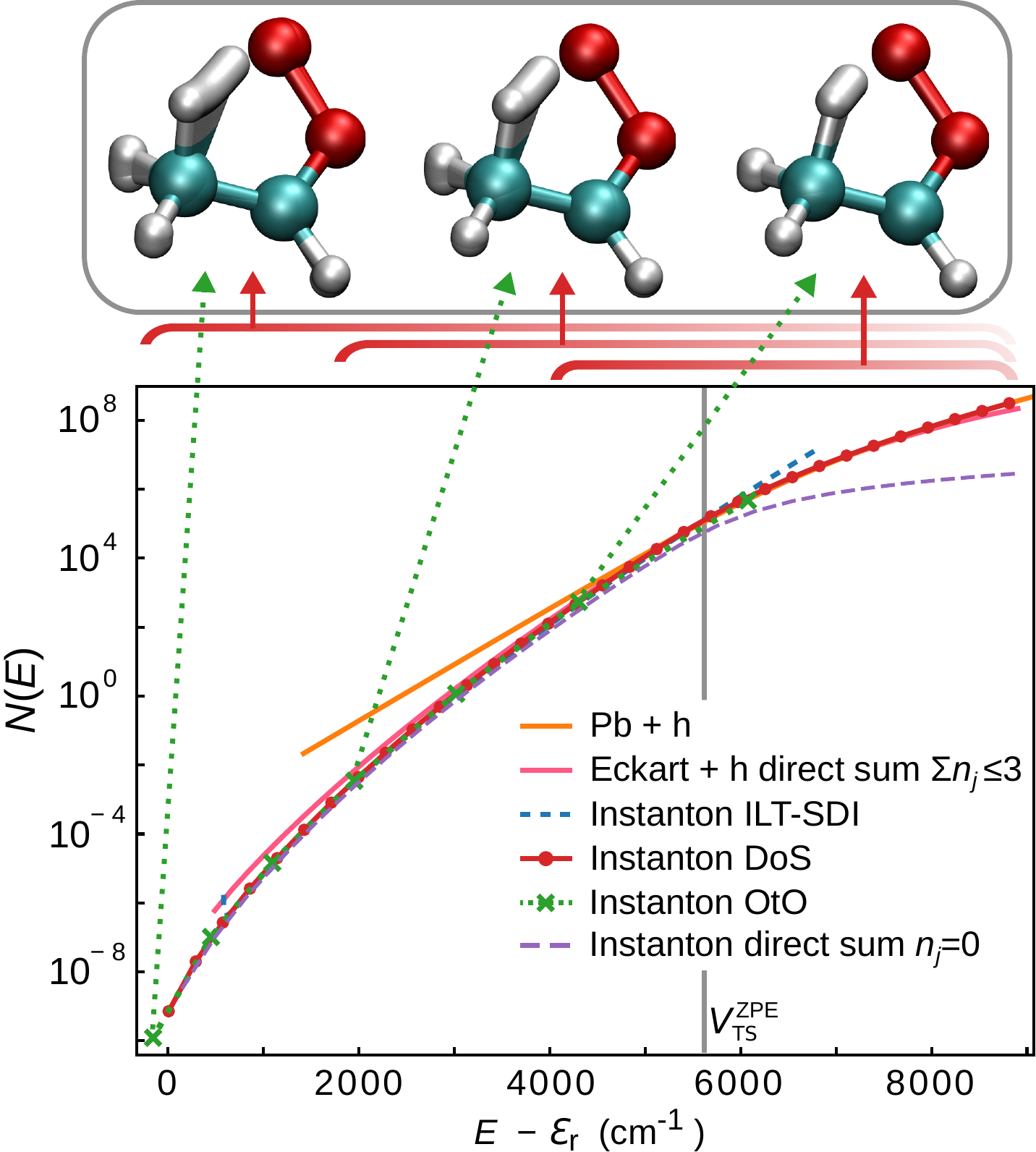}
    \caption{
    Cumulative reaction probability for the Criegee reaction, obtained using various methods.
    The top panel shows representative instanton configurations from left to right of deep, intermediate and shallow tunneling and 
    the green dotted arrows indicate the data points on the $N_\text{OtO}(E)$ curve to which they correspond.
    The red brackets show the range of the $N(E)$ curve (obtained with the DoS method) to which each instanton contributes, and the color gradient illustrates that this contribution diminishes as $E$ increases.
    $V^\text{ZPE}_\text{TS}$ for this reaction was calculated to be $5641 \, \mathrm{cm^{-1}}$.
    }
    \label{CriegeeNE}
\end{figure}

Images of instantons at three different energies are shown in Fig.~\ref{CriegeeNE}. 
We specifically indicate the point on the $N_\text{OtO}(E)$ curve to which each instanton corresponds within the OtO method.
One can see that the longer tunneling pathways correspond to lower values of energy.
Within the DoS method, each instanton makes a contribution to $N_\text{DoS}(E)$ for all $E \ge E_\text{I}+\mathcal{E}_\text{I}$ in a manner that the contribution diminishes the higher one goes in energy.
The corresponding cumulative reaction probabilities computed by a variety of methods from these instantons are also plotted in Fig. \ref{CriegeeNE}.

For this system, the $N(E)$ curves predicted by the different instanton-based methods are in good agreement with each other.
Taking the DoS method as the benchmark
suggests that there are no particular problems with employing the simpler 
ILT-SDI or OtO methods in this case.
Similarly to the H + H$_2$ reaction, the results appear to be reliable for energies below the barrier top, for the same reasons as discussed previously. 

Regarding the methods based on the separable approximation, the ``pb + h'' method seems reliable for the classical transmission above the barrier.
However, it clearly fails for the deep-tunneling regime because the true barrier does not resemble a parabola except near the top.
Interestingly however, the ``Eckart + h'' method works well for this system, with only a small deviation from the non-separable instanton methods even in the low energy deep-tunneling regime.
We found that it is the separable approximation that caused the slight deviation, rather than the approximation of the tunneling pathway itself. 
As we show in the SI, the abbreviated actions, $W(E_\text{I})$, computed from an Eckart model agree almost perfectly with that computed along the instanton, suggesting that this reaction can be well described with the Eckart model, and that corner-cutting effects in the tunneling pathways are small.
Since the Eckart barrier is such a good model for the barrier shape of this reaction, the ``MEP + h'' method is not shown as this is practically equivalent.
The similarity between ``Eckart + h'' and instanton theory (as has also been observed for similar reactions in \Ref{Watson2018Criegee}) explains why the simpler Eckart approach has been successfully used in many previous studies of tunneling.
However, although we believe that some reactions can be well described with the Eckart model, 
there will of course be systems for which it is not applicable and for which the instanton methodology is necessary.

Another feature of this system worth noting is the close agreement between the instanton method based only on the vibrational ground state (using direct summation over the rotational states only) and all the other instanton-based methods which take into account all rovibrationally excited states.
This implies that for energies below the barrier top, vibrational excitations do not contribute much to the tunneling process.
For transmission above the barrier, however, contributions from vibrationally excited states become very important and ignoring them could lead to orders of magnitude of error.
Because of this, among all the computationally feasible instanton-based methods tested here, only the DoS method has a reasonable prediction for energies above the barrier and connects smoothly with the ``pb + h'' result, which is expected to be accurate in this classical regime.

\begin{figure}[!ht]
    \centering
    \includegraphics[width=9cm]{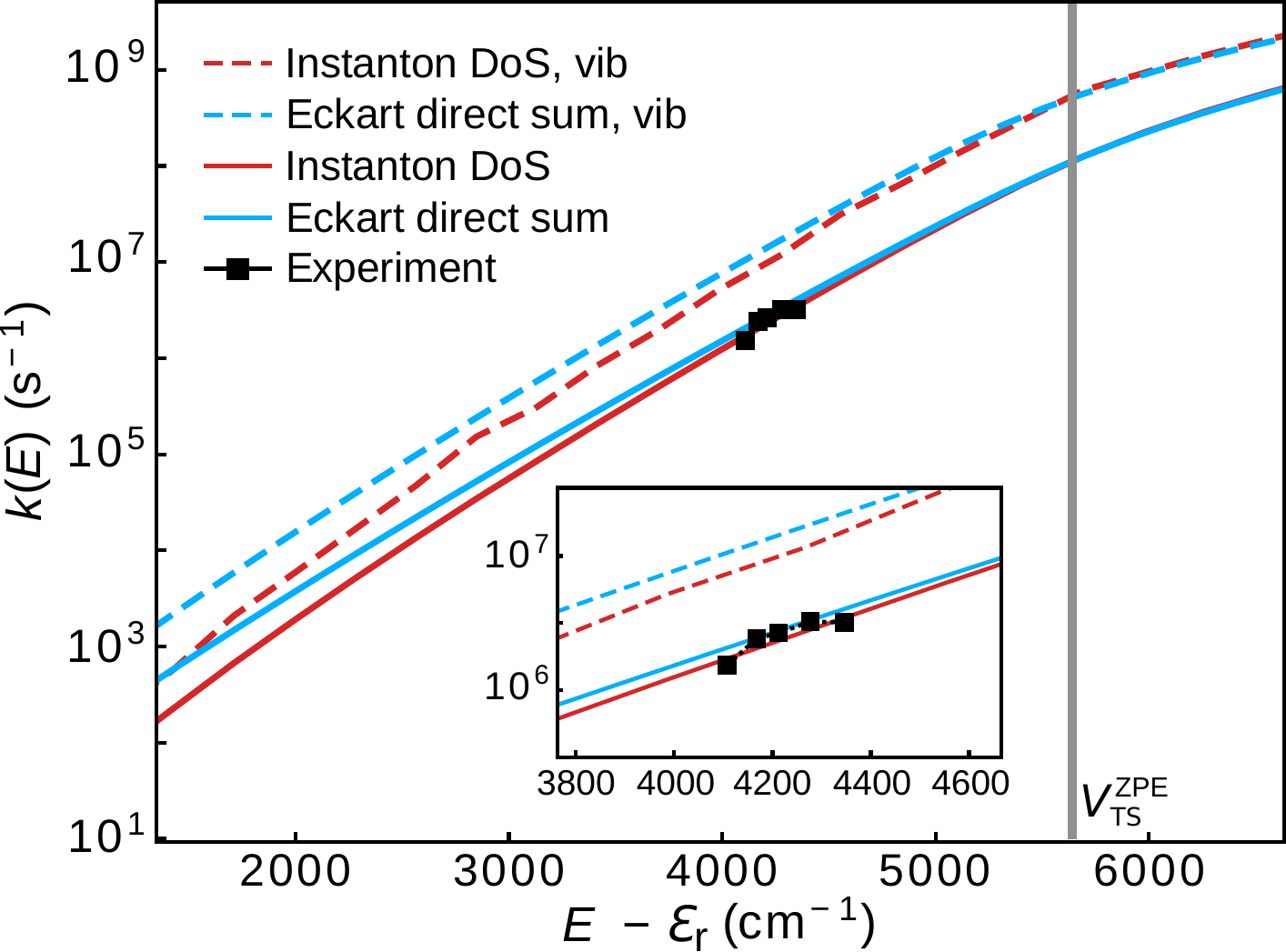}
    \caption{
    Microcanonical rate constants obtained with various methods for the unimolecular hydrogen transfer of \textit{syn-}\ch{CH3CHOO}.
    Rate constants which only include the vibrational states and neglect rotational effects are labeled with `vib'.
    The experimental rate constants from Ref.~\citenum{Fang2016CH3CHOO} are shown for comparison and the inset shows a zoom-in of the plot around the experimental data.
    }
    \label{Criegee_rate}
\end{figure}

For this reason, we therefore used $N_\text{DoS}(E)$ in the RRKM formulation [Eq.~\eqref{kMI}] to obtain microcanonical rate constants, $k(E)$, which are plotted against experimental results in Fig.~\ref{Criegee_rate}.
In order to compute the microcanonical rates, the reactant density of states, $\rho_\text{r}(E)$, is also needed, and we calculate it using the SPA-ILT of the reactant partition function, using an approach equivalent to Eq.~\eqref{ILTsp}.
When including all rotational and vibrational excitations in the cumulative reaction probability (i.e.\ using both rotational and vibrational partition functions in Eq.~\eqref{ILTsp}), the DoS method predicts microcanonical rate constants in excellent agreement with the experimental measurements of Ref.~\citenum{Fang2016CH3CHOO}.

We also examined the importance of rotational excitations on the microcanonical rate, by comparing with a calculation obtained directly from $\rho_\text{I}^\text{vib}(E;E_\text{I})$ and $\rho_\text{r}^\text{vib}(E)$, i.e.\ by excluding the rotational partition function in the ILT\@.
The results show that if rotational contributions were excluded, the predicted microcanonical rate would have been about four times higher than the experimental results. 
Although a factor of four is relatively small in the scope of rate constants, this suggests that the redistribution of the rotational energy whilst tunneling plays a non-negligible role in this reaction.
It is therefore important to include rotations in computing microcanonical rates if quantitative accuracy is desired.
We also note that when rotations are excluded, the $k(E)$ curve from the DoS method becomes slightly jagged.
Such behavior is expected due to the discrete nature of $\rho_\text{I}^\text{vib}(E;E_\text{I})$ at low energies near $E_\text{I}+\mathcal{E}_\text{I}$ (see Appendix A), suggesting that the steepest-descent approximation to the ILT can have minor numerical issues for small systems without rotations or soft vibrational modes.
The DoS method is designed for molecular systems with a high density of states, which may arise either from closely-spaced rotational energy levels or from a large number of vibrational modes, especially those with small frequencies.
These are also cases where the direct-sum approach will quickly become numerically impractical even at low energies, thus making the new method a good complement to the direct-sum approach.
%

In Ref.~\citenum{Fang2016CH3CHOO}, 
an Eckart approximation was used to explain the experimental results. 
Our calculations reproduced these previous results using the ``Eckart + h'' method. 
Note that it was again necessary to include the contribution of the rotational states, which was carried out using the conventional $J$-shifting approach.
A similarly good agreement with experiment could have been reached with the DoS instanton method
because in this case, the ``Eckart + h'' method is apparently a good approximation to it. 
In general, however, the instanton result is expected to be more reliable.
For instance, in this system we expect that at energies below 2000 cm$^{-1}$, the ``Eckart + h'' approximation would be less accurate and may overpredict the rates by a factor of 2 to 3.

\subsection{Bimolecular Reaction with a Pre-Reactive Complex}
In elementary bimolecular reactions, 
the mechanism is assumed to follow a reaction coordinate
along which 
the potential energy of the system increases monotonically as the reactant molecules approach each other until the transition state is reached and then decreases as the products are formed.
However, in many systems which exhibit strong intermolecular forces, this assumption is invalid and the reactant molecules first form a metastable complex with potential energy lower than that of the separated reactants.   From here, they may either react to form products or dissociate back to the reactant state.
Many reactions which exhibit pre-reactive complexes are of high interest for modeling kinetic networks in astrochemistry and atmospheric chemistry. \cite{multiwell2000,MESMER}
Due to the importance of nuclear quantum effects in such environments, instanton theory has been applied to similar systems in the past. \cite{Lamberts2017HH2S, Meisner2016review, Meisner2016water,Kastner2016Faraday} 
Some of these applications exhibit the characteristic breakdown of thermal instanton theory (SDI) at low temperatures and we seek to improve on this with the TMI rate computed from the DoS approach.

\begin{figure}[!ht]
    \centering
    \includegraphics[width=9.5cm]{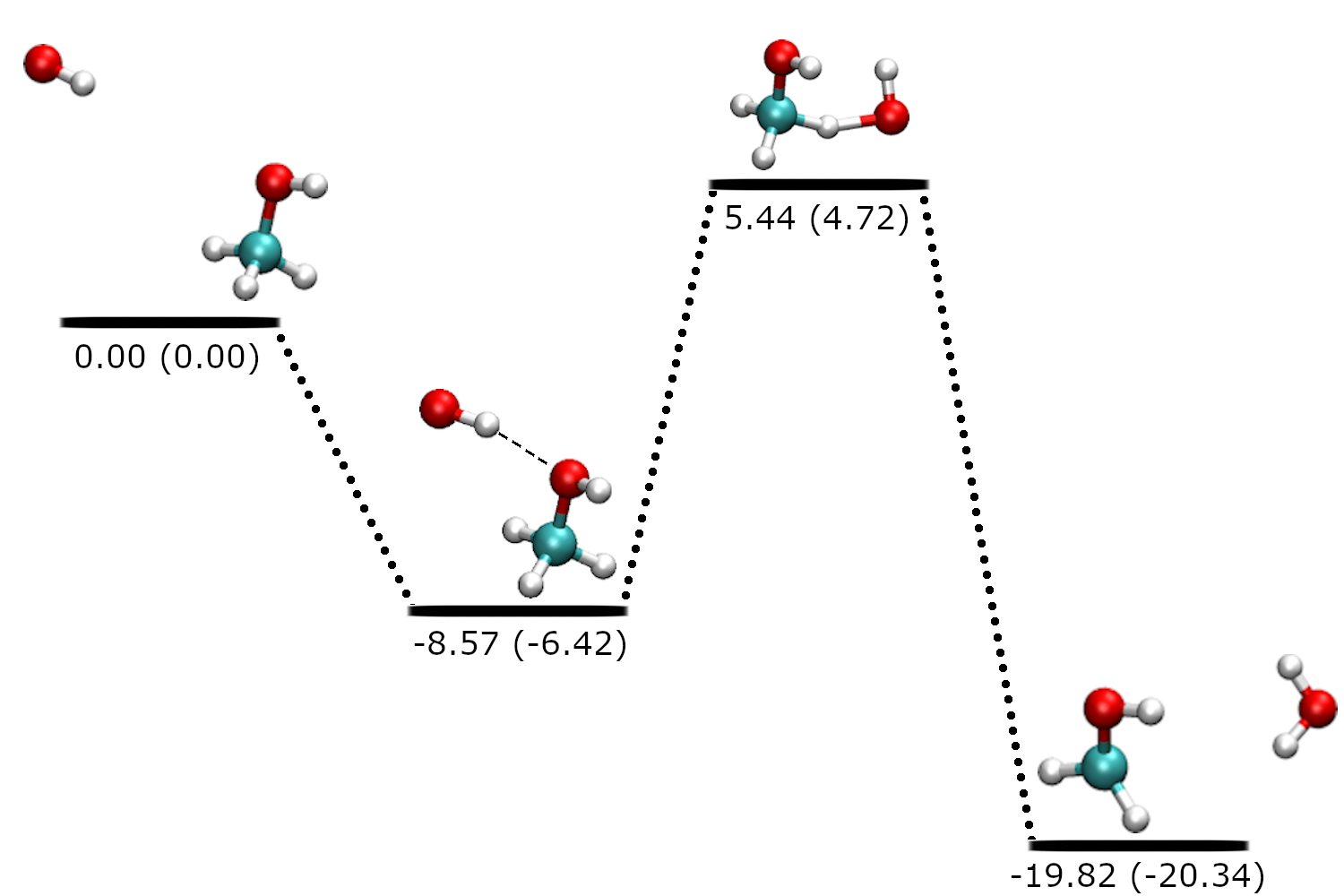}
    \caption{
    Reaction diagram for the \ch{CH3OH} + OH reaction studied in this work.
    Energy units are in kcal/mol and the numbers in parentheses correspond to ZPE-corrected energies relative to the separated reactants.
    All \textit{ab initio} calculations were carried out at the level of MP2/cc-pVDZ.
    }
    \label{rxn_diagram}
\end{figure}

For instance, in the hydrogen abstraction reaction \ch{CH3OH + OH -> [CH3OH ··· OH] -> CH2OH + H2O} depicted in Fig.~\ref{rxn_diagram}, the \ch{OH} radical forms a pre-reactive complex by hydrogen-bonding with the hydroxyl group of the methanol (\ch{CH3OH}) before abstracting a methyl hydrogen via a tunneling mechanism to form products. \cite{MethanolOH_Heard, MethanolOH_Cernicharo, MethanolOH_Truhlar}
Because the transition state is chiral but the reactants are not, $\sigma=2$ for this reaction.
There is also a secondary reaction, which we do not consider here, in which the hydroxyl hydrogen is abstracted instead. 
This secondary reaction has a pathway with only a slightly higher potential-energy barrier and should thus be included to study the total rate of loss of OH\@.  However, obtaining a local PES for this mechanism would require multi-reference electronic structure calculations
due to the proximity of the two oxygen atoms. \cite{MethanolOH_Truhlar}
We therefore choose to concentrate on the theoretical aspects of the barrier-crossing step of the primary reaction pathway and all our results refer to the rate constant of this process only.
The low-pressure limit of the reaction is considered here, 
which implies that we should not assume thermalization in the pre-reactive complex and the reactant partition function is given by the separated reactant molecules.
\Ref{MethanolOH_Truhlar} has carefully discussed how the entire reaction can be simulated in both the low- and high-pressure limits including the capture rate for the formation of the complex and its dissociation
in order to connect more directly to the scenario measured in experiment. \cite{MethanolOH_Heard, MethanolOH_Cernicharo}

\begin{figure}[!ht]
    \centering
    \includegraphics[width=9.5cm]{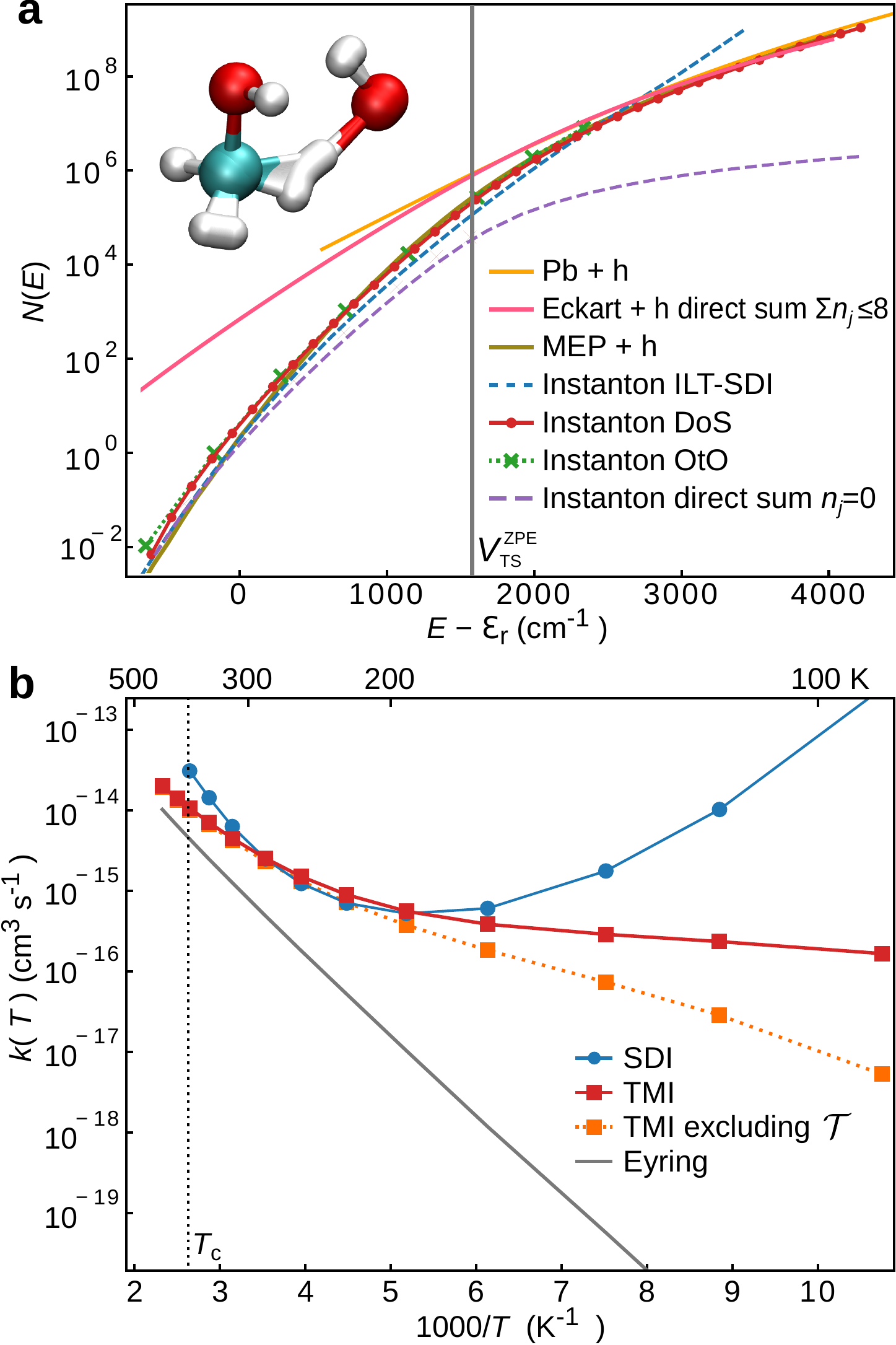}
    \caption{
    (a) Cumulative reaction probability for the \ch{CH3OH + OH -> CH2OH + H2O} reaction obtained using various methods.
    (b) Arrhenius plot of the thermal rate constant in the low-pressure limit computed with SDI [Eq.~\eqref{kSDI}], TMI [Eq.~\eqref{kint}], and Eyring TST [Eq.~\eqref{kEyring}] for comparison.
    }
    \label{Methanol}
\end{figure}

For the \ch{CH3OH} + OH reaction presented above, we calculate a Born--Oppenheimer PES on the fly from MP2/cc-pVDZ energies, gradients, and Hessians computed in Molpro.\cite{Molpro}
The results based on MP2 calculations however are not expected to be quantitatively accurate.  In fact, it was shown in a previous theoretical study that the transition-state energy computed with CCSD(T) is 1.46 kcal mol$^{-1}$,\cite{MethanolOH_Truhlar} which is considerably lower than that predicted by MP2.
Therefore our focus here is on showing that the DoS approach is qualitatively correct for a system with a pre-reactive complex and that it is stable enough to be used for typical systems in which the PES is calculated on the fly.
In other work, we have shown that it is possible to apply the instanton method quantitatively using machine-learning to dramatically reduce the number of high-level electronic-structure calculations required without affecting the accuracy of the result. \cite{GPR,*Muonium}
Eight instantons, each with 64 ring-polymer beads, were optimized in the temperature range 163 K to 378 K such that they were roughly evenly spaced in energy. \cite{PierreThesis}
The cumulative reaction probabilities calculated from these instantons are plotted in Fig.~\ref{Methanol} along with an Arrhenius plot of the thermal rate constants.
Additionally, three instantons with 128 beads were optimized at 133 K, 113 K, and 93 K for thermal (SDI) rate calculations.

In Fig.~\ref{Methanol}a, we compare the different methods for calculating the cumulative reaction probability, $N(E)$.
We obtain reasonable results from the DoS instanton method, which we assume to be the most reliable method tested here.  In particular, it matches with the ``pb + h'' method at high energies. 
In this system, we again see that the OtO method predicts $N(E)$ in excellent agreement with the DoS instanton method, although it is not available at high energies.
The ILT-SDI method is also in reasonably good agreement with the DoS instanton method, although it breaks down at high energies, in a similar fashion to the other systems studied in this work.

To understand the importance of vibrational excitations in the cumulative reaction probability, we computed $N_\text{SC-sum}(E)$ using only the ground-state vibrational excitations of the instantons (and also accounted for the rotationally excited states using the $J$-shifting approximation).
This obviously causes a dramatic underestimation of $N(E)$ at high energies.
However, in contrast to the previous two systems, even for energies below $V_{\text{TS}}^\text{ZPE}$, neglecting the vibrational excitations leads to a significant underestimation of $N(E)$ by almost an order of magnitude.
This is likely caused by the presence of multiple soft vibrational modes in this reaction process.
In fact, one needs to sum over all combinations of excited states up to a total excitation number of $\sum_j n_j=8$ to converge the direct summation within the ``Eckart + h'' approximation at the highest energies computed here, such that the summation has over one million terms, even for this small system of only eight atoms.
This means that even after the instantons are computed, the computational effort involved in post-processing of the direct summation approach can be prohibitively expensive, and would soon become impossible if more excited states were needed or if larger molecules were studied.
In contrast, the DoS approach remains computationally efficient and does not scale with system size.
The ``pb + h'' approximation to $N(E)$ breaks down, not only for low energies in the deep-tunneling regime, but also for energies near $V_{\text{TS}}^\text{ZPE}$.
Moreover, unlike in the previous cases, modifying the parabolic-barrier model to an asymmetric Eckart barrier barely improves the performance, suggesting that the real tunneling pathway is drastically different from either of these simple models.
The ``MEP + h'' approach appears to have resolved the problems of the Eckart model, predicting a $N(E)$ curve close to the instanton results.
However, further analysis shows that good agreement between the ``MEP + h'' model and the instanton results is deceptive, and we argue that this is in fact a coincidence due to error cancellation.
We show in the SI that the system has strong corner-cutting effects in the tunneling process, such that the abbreviated action, $W(E_\text{I})$, computed from the MEP is significantly higher than that of the instantons, which results in orders of magnitude of underestimation in $P_\text{SC}(E_\text{I})$ at low energies.
This error is however offset by another error introduced by the separable approximation, which neglects the frequency changes in the perpendicular modes for the long tunneling pathways.
Since one cannot expect that this error cancellation will occur in general, the ``MEP + h'' method should not be trusted for reactions with relatively complex pathways, such as this one whose mechanism involves large-amplitude motion of the OH molecule from its position in the pre-reactive complex before the H-bond can break (see Fig.~\ref{rxn_diagram}).
Instanton-based methods on the other hand have a clear advantage for these complex systems, especially in the deep-tunneling regime, as they can reliably account for corner-cutting effects for long tunneling pathways and do not rely on the separable approximation.

In Fig.~\ref{Methanol}b, we show results for thermal rate constants
and note that TMI tends to Eyring TST at high temperature in accordance with the quantum-classical correspondence principle. 
SDI, on the other hand, is known to overpredict the rate near 
the crossover temperature, $T_\mathrm{c}$, and is not applicable at higher temperatures.
However, the differences between TMI and SDI are much more dramatic at low temperature.
This is in spite of the fact that the microcanonical ILT-SDI method appears to be able to describe the cumulative reaction probability reasonably well.
In this case it is clear that the thermal SDI approach breaks down because it effectively ignores the lower limit of the energy integral and may include instanton tunneling  contributions with total energy below that of the separated reactants, $\mathcal{E}_\text{r}$, which is of course forbidden by energy conservation principles.
For example, the steepest-descent procedure selects instantons with negative energies ($E_\text{I}<0$) once the temperature is lowered below about 225 K\@.
This leads to an unphysical temperature-dependence of the forward rate constant \footnote{Note that there are situations in which the rate may increase at low temperature, but these are due to the competition of multiple processes\cite{Mozurkewich1984negative} which we do not consider here.}, a
problem which has also been observed in previous studies \cite{Meisner2016water,Kastner2016Faraday}.
The SDI method (with the appropriate reactant partition function) would however give reasonable predictions for the thermal rate in the high-pressure limit, in which the assumption that the pre-reactive complex is in thermal equilibrium (due to collisions with the surrounding gas) would be valid.

TMI provides a physically reasonable prediction of the low-pressure thermal rate constant by limiting the integration range over $E$ to energies above $\mathcal{E}_\text{r}$.
However, this approach is not equivalent to simply selecting instantons with
$E_\text{I}+\mathcal{E}_\text{I}>E_\text{r}$,
which would effectively neglect the $\mathcal{T}$ term in Eq.~(\ref{TMI_derive}c).
In fact, even sub-energetic instantons 
have excited parts of their density of states which lie above the reactant energy and should thus contribute to the rate.
These instantons are included in the $\mathcal{T}$ term (as explained in Sec.~2.5)
and cannot be neglected without significantly affecting the prediction at low temperatures.
As shown in Fig.~\ref{Methanol}b, $\mathcal{T}$ is important at low temperatures and even becomes the dominant term below about 170 K in this example.
This suggests that tunneling through the excited states of long tunneling pathways (which are at low energies) is crucial at low temperature.

Summarizing the results for this system, we see that the instanton approaches predict a large tunneling effect at low temperatures,
but only the TMI method is expected to be reliable.
We therefore expect that 
this approach
will be a viable way to model typical atmospheric and astrochemical reaction rate constants for large molecular systems with pre-reactive complexes.

\section{Conclusion}
We have proposed an effective and efficient formula to compute the microcanonical cumulative reaction probability from density-of-states instanton calculations, using the steepest-descent approximation to the inverse Laplace transform.
This approach 
has made it feasible to include tunneling effects into 
RRKM theory without employing the separable approximation and without requiring the computationally intensive task of summing over all the quantum states.
The new method can also be used to accurately predict thermal rate constants in molecular reactions for which conventional instanton approaches are not valid, such as systems with pre-reactive complexes.

We have tested the various approaches on three different molecular systems: a bimolecular H + \ch{H2} scattering reaction 
to benchmark against exact results;
a unimolecular hydrogen transfer in photoexcited \textit{syn-}\ch{CH3CHOO} 
which is characterized by a microcanonical rate constant;
and an on-the-fly \textit{ab initio} calculation of the thermal rate of \ch{CH3OH} + OH to demonstrate that we can treat pre-reactive complexes in a typical simulation environment. 
We find that the new DoS instanton method is accurate and efficient in all cases.
We also found that our simpler OtO method agrees well with this new method, for all the systems tested, suggesting that in many cases it could be the easiest way to predict the microcanonical rate, at least at energies below the barrier, where it is applicable. 
However, we do expect that in certain cases it will be less reliable than the DoS method, such as for reactions with particular barrier shapes.\cite{broadtop}
Furthermore, we demonstrated that instanton methods are typically more reliable compared to methods based on the separable approximation, such as the commonly used ``Eckart + h'' method, whose accuracy can be strongly system dependent.
The new approach will thus be useful in providing a practical computational method which can be used to replace these simpler approximations when they fail.

\section*{Appendix}
\appendix
\section{The Inverse Laplace Transform} \label{app:ILT}
The Laplace transform of $f(E)$ is defined as
\begin{equation}
F(\beta)=\mathcal{L}[f(E)](\beta)=\int_0^\infty \text{d}E~f(E)\mathrm{e}^{-\beta E}.
\end{equation}
Likewise, $f(E)$ can be uniquely determined by the inverse Laplace transform of $F(\beta)$,
\begin{equation}
f(E)=\mathcal{L}^{-1}[F(\beta)](E)=\frac{1}{2\pi i}\lim_{B\rightarrow\infty}\int_{\beta_0-iB}^{\beta_0+iB}\text{d}\beta~\mathrm{e}^{\beta E}F(\beta)
\end{equation}
where $\beta_0$ is a real number larger than the real part of all the singularities of $F(\beta)$.
For our applications, one can select $\beta_0\rightarrow0_+$, and make the Wick rotation $v=i\beta$,
\begin{equation}
f(E)=\frac{1}{2\pi}\int_{-\infty}^{\infty}\text{d}v~\mathrm{e}^{-ivE+\ln F(-iv)}.
\label{ILT_0p}
\end{equation}
Making the stationary-phase approximation to Eq.~\eqref{ILT_0p}, for a given $E$, the stationary point $\beta_\text{sp}$ satisfies
\begin{subequations}
\begin{align}
E=-i\frac{\partial \ln F(-iv)}{\partial v}=-\left(\frac{\partial \ln F(\beta)}{\partial \beta}\right)_{\beta=\beta_\text{sp}},
\end{align}
and at this value of $E$, the function is approximately
\begin{align}
f(E)=\left(2\pi\frac{\partial^2\ln F(\beta)}{\partial \beta^2}\right)_{\beta=\beta_\text{sp}}^{-\frac{1}{2}}\mathrm{e}^{\beta_\text{sp}E+\text{ln}F(\beta_\text{sp})}.
\end{align}
\label{SPA}
\end{subequations}

As an example, the rotational partition function of a nonlinear molecule is commonly approximated as 
$Z^\text{rot}(\beta)= \alpha\beta^{-3/2}$, where $\alpha=\sqrt{8\pi\det\textbf{I}/\hbar^6}$ and
\textbf{I} is the moment of inertia about the center of mass.
The exact ILT (i.e., the rotational density of states $\rho^\text{rot}(E)$) is simply
$\mathcal{L}^{-1}[Z^\text{rot}(\beta)](E)=\alpha\frac{2}{\sqrt{\pi}}\sqrt{E}$.
The stationary-phase approximation, however, gives
\begin{subequations}
\begin{equation}
E=-\left(\frac{\partial\ln Z^{\text{rot}}(\beta)}{\partial\beta}\right)_{\beta=\beta_{\text{sp}}}=\frac{3}{2}\frac{1}{\beta_\text{sp}},~~\beta_\text{sp}\in\mathbb{R}^+
\end{equation}
\begin{equation}
\mathcal{L}^{-1}[Z^\text{rot}(\beta)](E)\approx\left(2\pi\frac{3}{2}\frac{1}{\beta_\text{sp}^2}\right)^{-\frac{1}{2}}\mathrm{e}^\frac{3}{2}\alpha\beta_\text{sp}^{-\frac{3}{2}}=1.056\alpha\frac{2}{\sqrt{\pi}}\sqrt{E} ,
\end{equation}
\end{subequations}
which has only a small 5.6\% error in its prefactor.

Next we examine the vibrational partition function described with the harmonic oscillator model.
For a one dimensional harmonic oscillator with frequency $\omega$, the partition function is $Z^{\text{vib-1D}}(\beta)=[2\sinh(\beta\hbar\omega/2)]^{-1}$, and the vibrational density of states calculated with SPA-ILT is given by,
\begin{subequations}
\begin{equation}
E=-\left(\frac{\partial\ln Z^{\text{vib-1D}}(\beta)}{\partial\beta}\right)_{\beta=\beta_{\text{sp}}}=\frac{\hbar\omega}{2}\coth\left(\frac{\beta_\text{sp}\hbar\omega}{2}\right)
\end{equation}
\begin{equation}
\rho^\text{vib-1D}_\text{SPA}(E)=\frac{1}{\sqrt{2\pi}}\frac{1}{\hbar\omega}\left(\frac{2E+\hbar\omega}{2E-\hbar\omega}\right)^{\frac{E}{\hbar\omega}}h(E-\tfrac{1}{2}\hbar\omega),
\end{equation}
\end{subequations}
where $h(E)$ is the Heaviside function.
In the limit $\beta_\text{sp}\rightarrow\infty$, the energy tends to the ground state,  $E\rightarrow\frac{1}{2}\hbar\omega$, and hence $\rho^\text{vib-1D}_\text{SPA}(E)\rightarrow\infty$, which mimics the exact DoS ($\delta$-function) at the ground state.
In the large $E$ limit, the continuous DoS is $\frac{1}{\hbar\omega}$, and one finds that $\rho^\text{vib-1D}_\text{SPA}(E)\rightarrow\frac{\mathrm{e}}{\sqrt{2\pi}}\frac{1}{\hbar\omega}=1.084\frac{1}{\hbar\omega}$.
Therefore, except for small numerical factors, the DoS obtained from SPA-ILT has the correct behavior at the two limits.
For the multidimensional case ($f$ independent harmonic oscillators),
\begin{subequations}
\begin{equation}
E=-\left(\frac{\partial\ln Z^{\text{vib}}(\beta)}{\partial\beta}\right)_{\beta=\beta_{\text{sp}}}=\sum_{j=0}^{f-1}\frac{\hbar\omega_j}{2}\coth\left(\frac{\beta_\text{sp}\hbar\omega_j}{2}\right),
\end{equation}
and
\begin{equation}
\rho^\text{vib}_\text{SPA}(E(\beta_\text{sp}))=\frac{1}{\sqrt{2\pi}}\left[\sum_{j=0}^{f-1}\left(\frac{\hbar\omega}{2}\right)^2\cosech^2\left(\frac{\beta_\text{sp}\hbar\omega}{2}\right)\right]^{-\frac{1}{2}}\mathrm{e}^{\beta_\text{sp} E(\beta_\text{sp})}Z^\text{vib}(\beta_\text{sp}).
\end{equation}
\end{subequations}
In the ground state limit ($\beta_\text{sp}\rightarrow\infty$ and $E\rightarrow\sum_{j=0}^{f-1}\frac{1}{2}\hbar\omega_j$), we again find that the density of states behaves like a delta function,  $\rho^\text{vib}_\text{SPA}(E)\rightarrow\infty$.
In the high-energy limit, $\beta_\text{sp}\rightarrow0^+$, $E\rightarrow f/\beta_\text{sp}$, one finds,
\begin{equation}
\rho^\text{vib}_\text{SPA}(E)\rightarrow \frac{1}{\sqrt{2\pi}}\sqrt{f}\left(\frac{\mathrm{e}}{f}\right)^{f}E^{f-1}\prod_{j=0}^{f-1}\frac{1}{\hbar\omega_j}\approx\frac{E^{f-1}}{(f-1)!\prod_{j=0}^{f-1}\hbar\omega_j},
\end{equation}
using the relation $f!\approx\sqrt{2\pi f}\left(\frac{f}{e}\right)^f$.
This shows that it recovers the well-known continuum approximation (also known as the Thomas--Fermi approximation) to the density of states of $f$ quantum harmonic oscillators when $E$ is much larger than $\hbar\omega_j$.
%

\begin{acknowledgement}

The authors acknowledge financial support from the Swiss National Science Foundation through Project 175696.

\end{acknowledgement}

\begin{suppinfo}

Further details of the simulations and additional discussion on the results.

\end{suppinfo}

\providecommand{\latin}[1]{#1}
\makeatletter
\providecommand{\doi}
  {\begingroup\let\do\@makeother\dospecials
  \catcode`\{=1 \catcode`\}=2 \doi@aux}
\providecommand{\doi@aux}[1]{\endgroup\texttt{#1}}
\makeatother
\providecommand*\mcitethebibliography{\thebibliography}
\csname @ifundefined\endcsname{endmcitethebibliography}
  {\let\endmcitethebibliography\endthebibliography}{}

\section{Supplementary data and discussions for the H + H$_2$ reaction}
\begin{figure}[!ht]
    \centering
    \includegraphics[width=8cm]{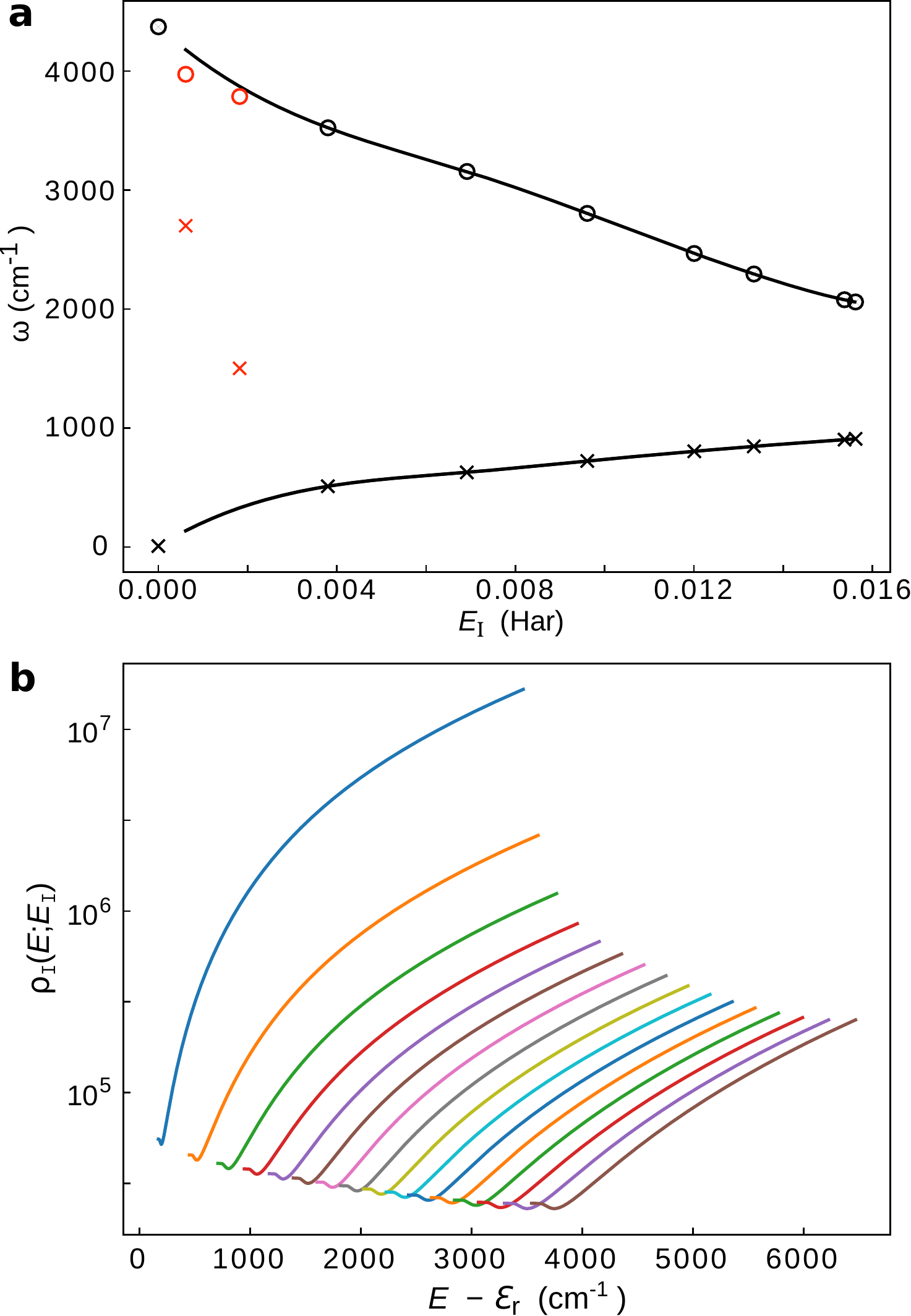}
    \caption{
    (a) Instanton fluctuation frequencies obtained from stability parameters of the instantons, marked by `x' and `o'.
    The vibrational mode labeled `x' corresponds to the bending mode and is two-fold degenerate, whereas `o' is the symmetric stretch.
    The red symbols mark the numerically unstable stability parameters from instantons with long imaginary times.
    The black curves are obtained from spline interpolations of the numerically stable stability parameters and the asymptotic result at $E_\text{I}=0$.
    (b) $\rho_\text{I}(E;E_\text{I})$ for the H+H$_2$ reaction obtained from the stationary-phase approximation to the inverse Laplace transform at different $E_\text{I}$ values.
    The starting point of each colored line corresponds to the instanton energy plus the zero point energy (ZPE) of the instanton ($E=E_\text{I}+\mathcal{E}_\text{I}(E_\text{I})$).
    }
    \label{HH2_SI}
\end{figure}
Fig.~\ref{HH2_SI}a shows that as the instanton path becomes longer (or as the instanton energy $E_\text{I}$ decreases), the instanton fluctuation corresponding to the H--H--H bending mode becomes smaller and the instanton fluctuation corresponding to stretching mode becomes larger.
For long imaginary time instanton paths, the stability parameters corresponding to the low frequency modes cannot be computed numerically with existing methods without introducing further approximations.
Fig.~\ref{HH2_SI}b shows the density of states of instantons.
At low $E_\text{I}$, due to the existence of low frequency bending modes, the density of states curve is particularly steep. 

\begin{figure}[!ht]
    \centering
    \includegraphics[width=9.5cm]{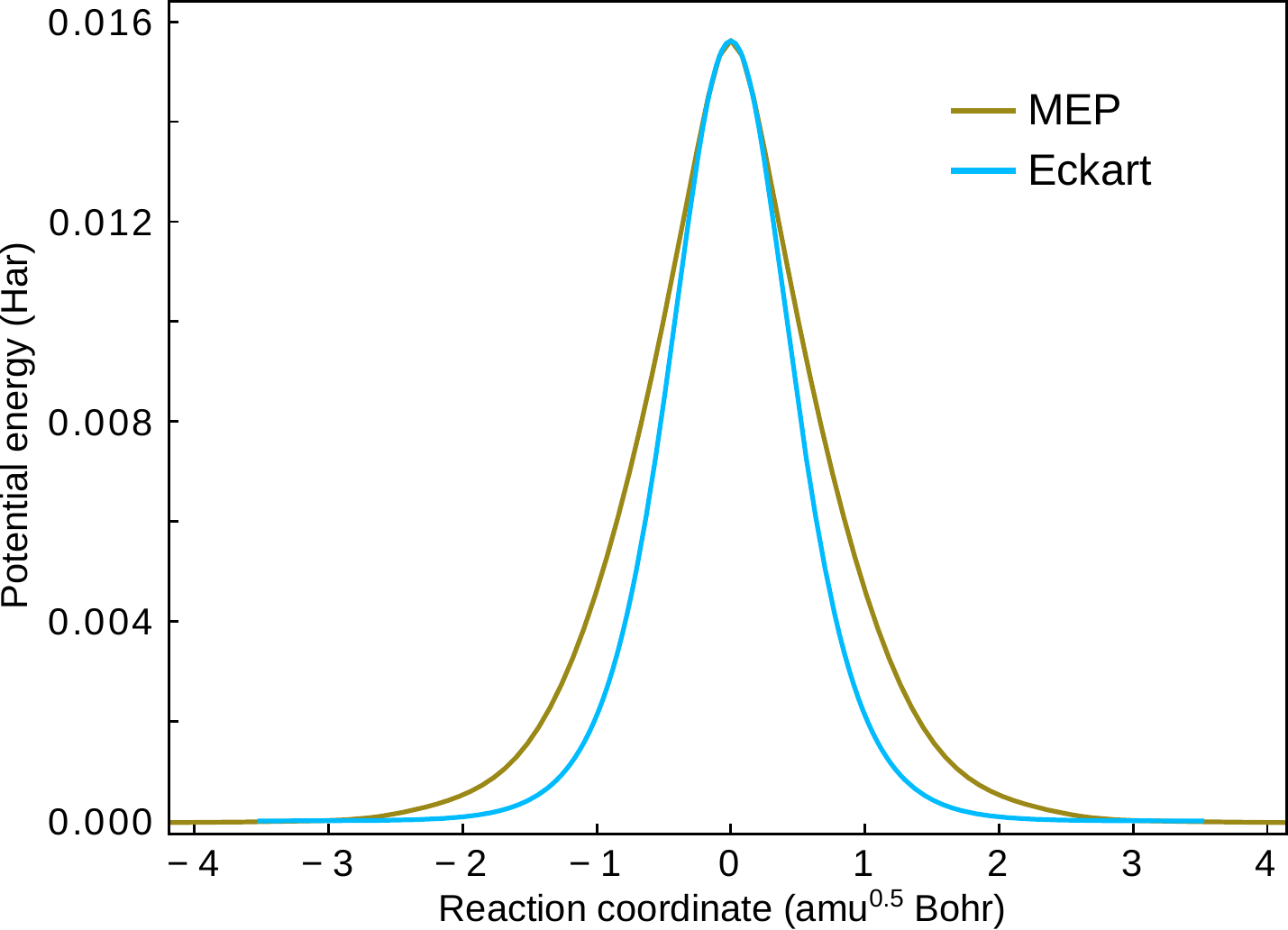}
    \caption{
    Comparison of the minimal energy pathway (MEP) and the Eckart model for the H + H$_2$ reaction.
    }
    \label{HH2_barrier}
\end{figure}
Fig.~\ref{HH2_barrier} shows that except at near the barrier top, the minimal energy pathway (MEP) is considerably broader than the Eckart model.
This is the main reason why the Eckart barrier overpredicts tunneling at low energies for this reaction. 

\begin{figure}[!ht]
    \centering
    \includegraphics[width=17cm]{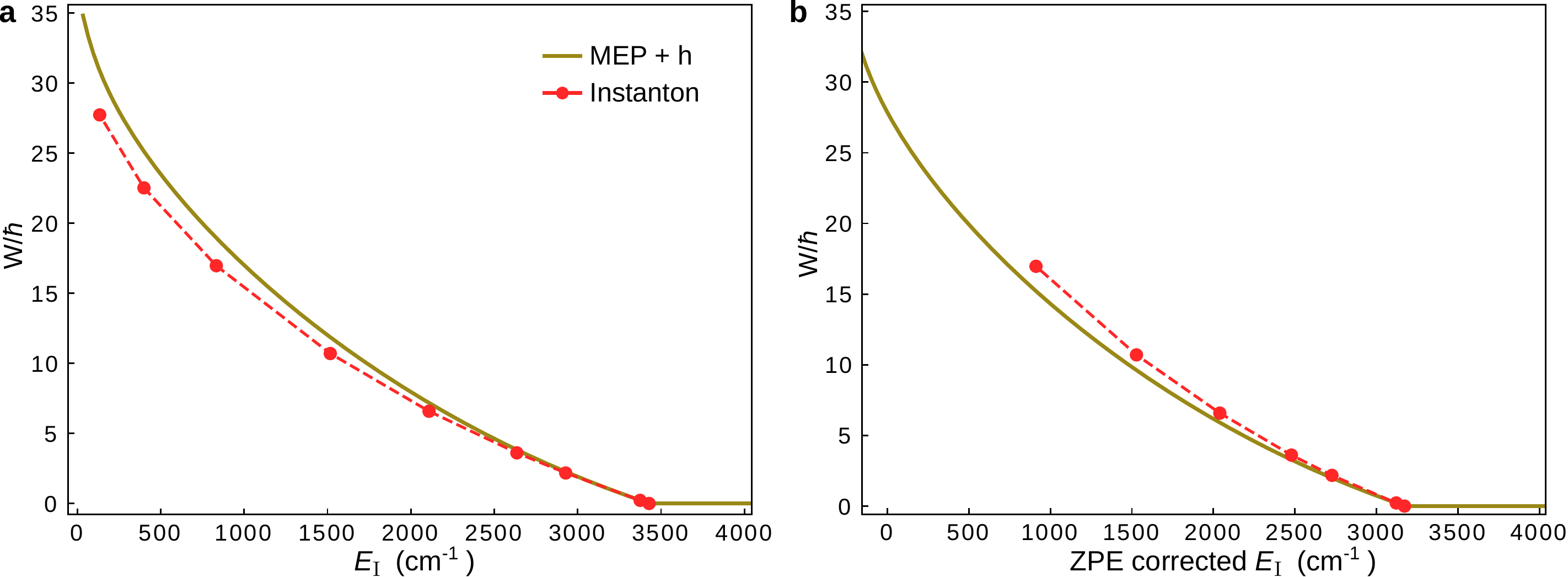}
    \caption{
    Comparison of the abbreviated action computed along the MEP and along the instanton for the H + H$_2$ reaction, plotted against (a) the instanton energy $E_\text{I}$, and (b) the ZPE corrected instanton energy,
    defined as $E_\text{I}+\mathcal{E}_\text{I}(E_\text{I})-\mathcal{E}_\text{r}$ for the instanton, and $E_\text{I}+\mathcal{E}_\text{TS}^{(0)}-\mathcal{E}_\text{r}$ for the ``Eckart + h'' (separable) approximation.
    }
    \label{W_cmp_HH2}
\end{figure}
Fig.~\ref{W_cmp_HH2}a shows that the abbreviated action, $W$, computed along the MEP is slightly higher than that of the instantons.
This is caused by corner-cutting effects, where the optimal tunneling pathway deviates from the MEP\@.
However, the difference is not large, which implies that the corner-cutting effects in this reaction are rather small. 
Fig.~\ref{W_cmp_HH2}b shows that the ZPE of tunneling pathways at low energies is underestimated in the separable approximation, making them more ``accessible" compared to the instanton results.
This is why the $N(E)$ predicted by the ``MEP + h'' method is counter-intuitively higher than the instanton and exact results.
\clearpage

\section{Supplementary data and discussions for the Criegee intermediate}
\begin{figure}[!ht]
    \centering
    \includegraphics[width=17cm]{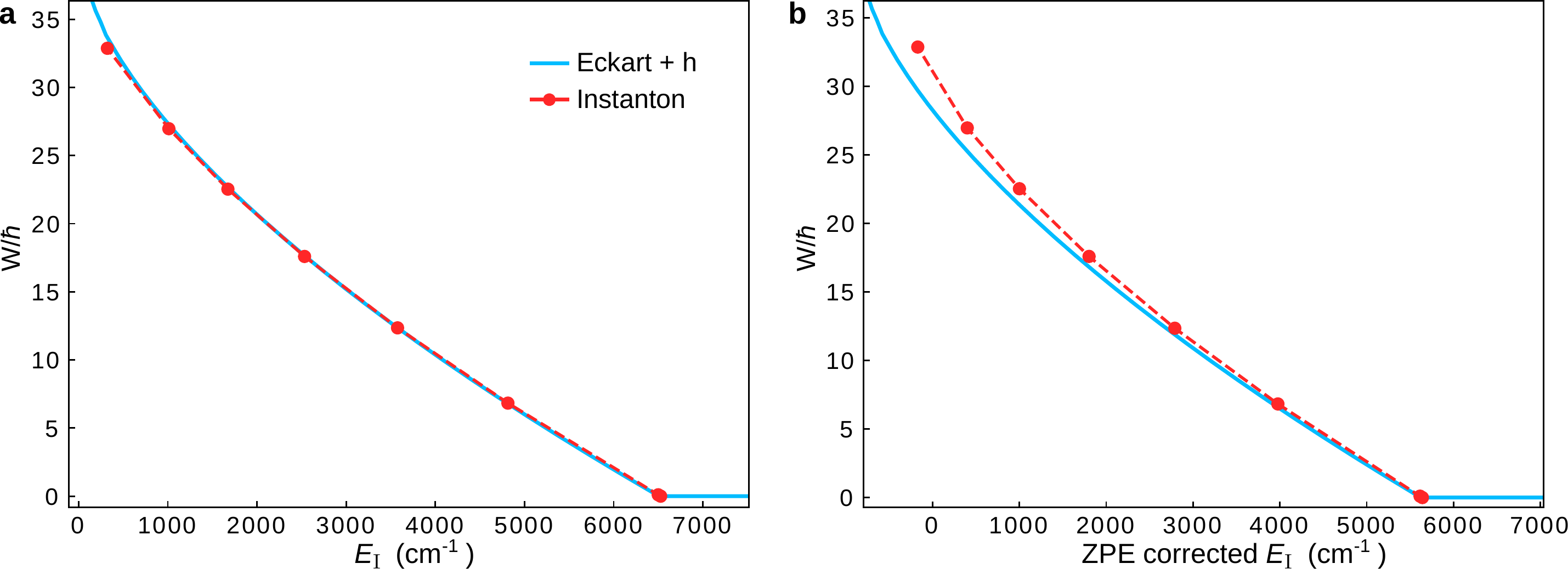}
    \caption{
    Comparison of the abbreviated action computed along the instanton and an effective abbreviated action computed on the Eckart barrier for the Criegee intermediate, plotted against (a) the instanton energy $E_\text{I}$, and (b) the ZPE corrected instanton energy,
    defined as $E_\text{I}+\mathcal{E}_\text{I}(E_\text{I})-\mathcal{E}_\text{r}$ for the instanton, and $E_\text{I}+\mathcal{E}_\text{TS}-\mathcal{E}_\text{r}$ for the ``Eckart + h'' (separable) approximation.
    The effective abbreviated action for the Eckart barrier is defined as $W(E_\text{I}) = \hbar\ln\left(\frac{1}{P_\text{Eckart}(E_\text{I})}-1\right)$, where $P_\text{Eckart}$ is the exact transmission probability for the Eckart barrier.
    }
    \label{W_cmp_criegee}
\end{figure}
Fig.~\ref{W_cmp_criegee}a shows that the abbreviated action computed on the Eckart barrier model agrees well with the abbreviated action obtained from the instanton theory for this system.
This means that the system is well described by the Eckart model, and suggests that corner-cutting effects are small here.
Fig.~\ref{W_cmp_criegee}b shows that it is the separable approximation that caused the slight overestimation of $N(E)$ compared to the instanton results
as it underestimates the ZPE of the longer tunneling pathways.
%
%
\clearpage

\section{Supplementary data and discussions for the Methanol + OH reaction}
\begin{figure}[!ht]
    \centering
    \includegraphics[width=17cm]{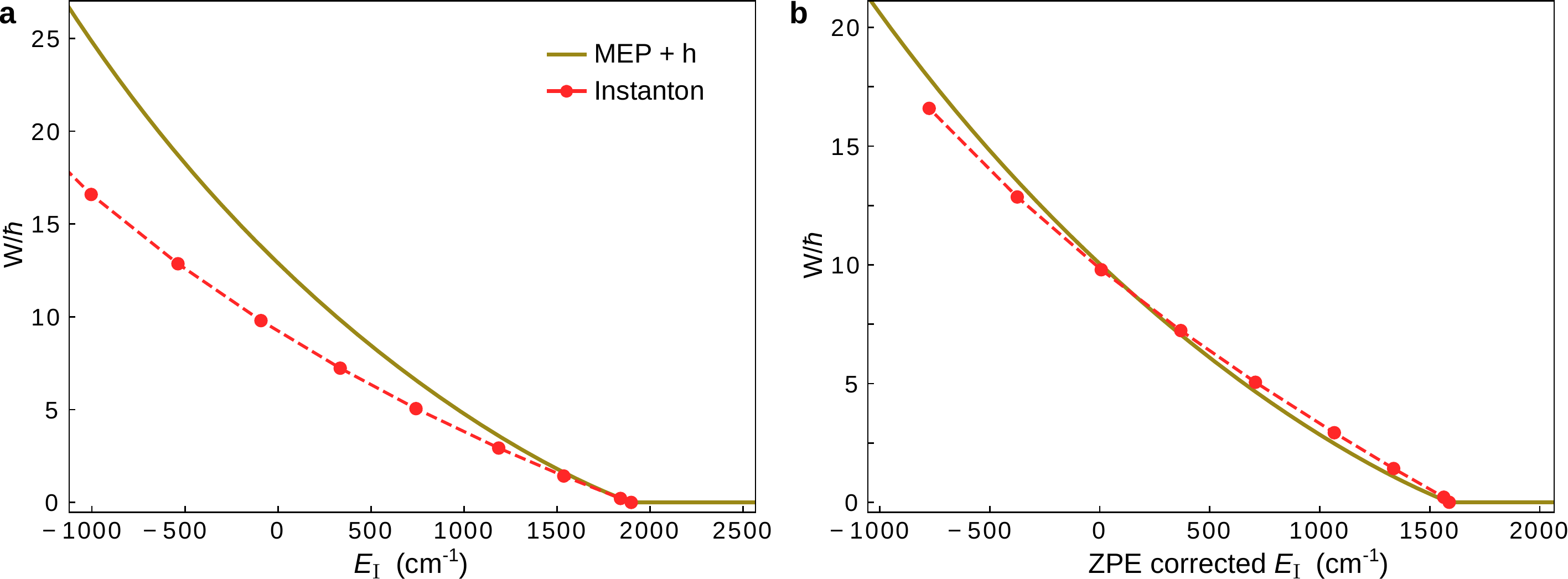}
    \caption{
    Comparison of the abbreviated action computed along the MEP vs along the instanton for the methanol + OH reaction, plotted against (a) the instanton energy $E_\text{I}$ ($E_\text{I}=0$ at the reactant), (b) the ZPE corrected instanton energy,
    defined as $E_\text{I}+\mathcal{E}_\text{I}(E_\text{I})-\mathcal{E}_\text{r}$ for the instanton, and $E_\text{I}+\mathcal{E}_\text{TS}-\mathcal{E}_\text{r}$ for the ``MEP + h'' (separable) approximation.
    }
    \label{W_cmp_meth}
\end{figure}
\begin{figure}[!ht]
    \centering
    \includegraphics[width=9cm]{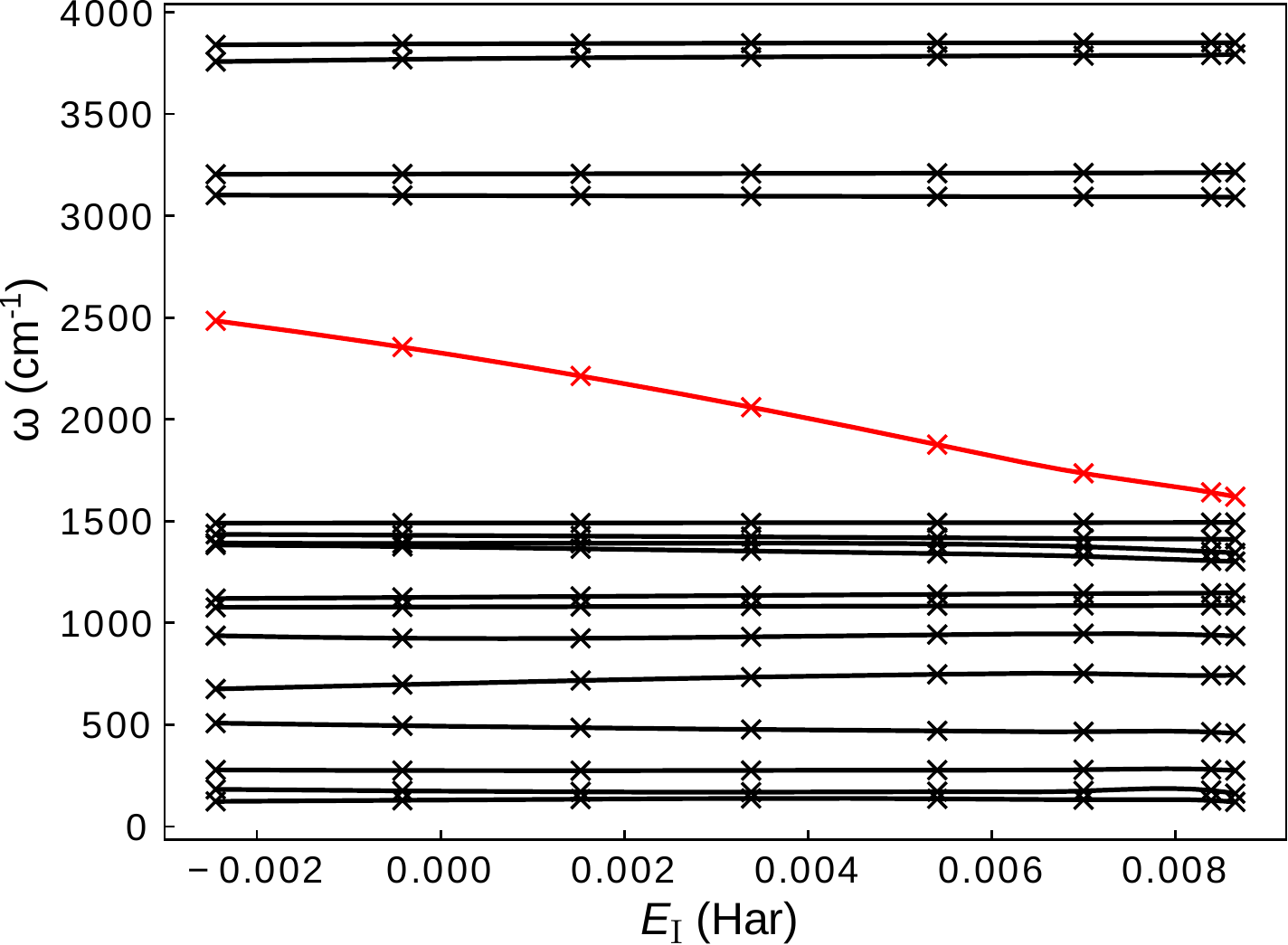}
    \caption{
    Instanton fluctuation frequencies for the methanol + OH reaction obtained from stability parameters of instanton at a range of energies.
    The red line highlights the mode whose frequency changes dramatically with $E_\text{I}$.
    }
    \label{stab_meth}
\end{figure}

Fig.~\ref{W_cmp_meth}a shows that the abbreviated action calculated along the MEP is severely overestimated compared to predictions from instanton theory.
This means that there are strong corner-cutting effects in this systems and that the MEP provides a poor description of the tunneling pathway.
Fig.~\ref{stab_meth} shows that among the instanton fluctuation frequencies obtained from the stability parameters, one frequency, which corresponds to the O--H--C bending mode, increases considerably when $E_\text{I}$ decreases.
This indicates that the corresponding mode is strongly coupled to the reaction degree of freedom and that the separable approximation breaks down.
Using the separable approximation will underestimate the effective ZPE of the long tunneling pathways. 
Fig.~\ref{W_cmp_meth}b shows that the error from the separable approximation cancels to a large extent with the error arising from neglecting corner cutting for this system.
Note that this large error cancellation should not be expected to happen for other reactions in general.

\end{document}